\documentclass[twocolumn,aps,nofootinbib,pre]{revtex4}

\usepackage{amsmath,amsfonts}
\usepackage{amssymb}
\usepackage{mathrsfs}

\newcommand{\be}{\begin{equation}}
\newcommand{\ee}{\end{equation}}

\newcommand{\E}{\mathcal{E}}
\newcommand{\ov}{\overline}
\newcommand{\ra}{\rangle}
\newcommand{\la}{\langle}

\newcommand{\T}{{\cal T}}

\newcommand{\ww}{\widetilde}

\newcommand{\cs}{\mathcal{S}}
\newcommand{\HH}{{\mathcal{H}}}

\newcommand{\D}{{\mathcal{D}}}
\newcommand{\N}{{\mathscr{N}}}
\newcommand{\R}{{\mathcal{R}}}
\newcommand{\tr}{{\rm Tr}}

\begin{document}

\title{
 Closeness of the reduced density matrix of an interacting small system to the Gibbs state
}

\author{Wen-ge Wang \footnote{ Electronic address: wgwang@ustc.edu.cn}}

\affiliation{ Department of Modern Physics, University of Science and Technology of China,
 Hefei 230026, China}
\date{\today}

\begin{abstract}
 I study the statistical description of a small quantum system,
 which is coupled to a large quantum environment in a generic form and with a generic interaction strength,
 when the total system lies in an equilibrium state described by a microcanonical ensemble.
 The focus is on the difference between the reduced density matrix (RDM) of the central system
 in this interacting case and the RDM obtained in the uncoupled case.
 In the eigenbasis of the central system's Hamiltonian,
 it is shown that the difference between diagonal elements is
 mainly confined by the ratio of the maximum width of the eigenfunctions of the total system in the uncoupled basis
 to the width of the microcanonical energy shell; meanwhile, the difference between off-diagonal elements
 is given by the ratio of certain property of the interaction Hamiltonian to the related level spacing
 of the central system.
 As an application, a sufficient condition is given, under which the RDM may have a canonical Gibbs form
 under system-environment interactions that are not necessarily weak;
 this Gibbs state usually includes certain averaged effect of the interaction.
 For central systems that interact locally with 
 many-body quantum chaotic systems, it is shown that the RDM usually has a Gibbs form.
 I also study the RDM which is computed from a typical state of the total system within an energy shell.
\end{abstract}

\maketitle

\section{Introduction}

\subsection{Motivations}

 In quantum statistical mechanics,
 one important topic is about the relationship between microcanonical(MC)-ensemble description
 and canonical-ensemble description.
 In particular, for a generic, isolated, and large quantum system that is described by an MC ensemble,
 the condition, under which the reduced density matrix (RDM) of an
 interacting small subsystem may have a canonical Gibbs form,  is still a problem not completely solved.
 Unlike the corresponding problem in the classical statistical mechanics,
 which can be solved relatively easily (see, e.g., Ref.~\cite{LL-SP}), this problem is highly-nontrivial,
 due to the mathematical difficulty met when dealing with the total  energy eigenstates under
 nonnegligible subsystem-environment interactions.

 A related important topic is justification of the usage of an MC-ensemble description
 for the total system, in view of the fact that quantum mechanics in principle allows a pure-state
 description for the total system.
 This topic is also of relevance, in the effort of generalizing equilibrium-state statistical-mechanics principles
 to nonequilibrium processes.
 Modern studies show that a mathematical concept
 related to high-dimensional linear space, namely typicality, plays an important role
 \cite{EisertFG15-NP,Tasaki16-typi-therm,GE16-thermal-review,Mori-IKU17-review},
 an idea that can be traced back to von Neumann's original work \cite{Qergo-Neuam}.
 Recently, by making use of the so-called Levy's lemma \cite{Levy-lemma,Linden2009pre},
 a quantitative progress was reported in  Ref.~\cite{PSW06} in 2006,
 wherein an upper bound is derived for the distance between the RDM computed from an MC-ensemble
 description  of the total system and that from a typical-state description of the total system.
 \footnote{Below, we refer to these two types of RDM as MC-ensemble-computed RDM and typical-state-computed RDM,
 respectively.}
 The derived upper bound shows that
 the MC and typical-state descriptions for the total system are effectively identical in view of
 computing the RDM of a small subsystem, when the dimension of the effective environmental state space
 is sufficiently large.

 In the same year of 2006, it was shown in Ref.~\cite{Goldstein06} that
 the RDM of a small subsystem, which is computed from a typical state of the total system,
 is typically close to the Gibbs state when the subsystem-environment interaction is very weak,
 under the well-known assumption about the exponential shape of the density of states of the environment.
 However, the strength of interaction required in the treatment of Ref.~\cite{Goldstein06} is usually too weak
 for a macroscopic environment to satisfy,
 due to the exponential increase of its density of states with the particle number \cite{RGE12}.
 To solve this problem, weak (not necessarily extremely weak) interaction was studied in Ref.~\cite{RGE12} in 2012,
 wherein an upper bound was given to the distance between two MC-ensemble-computed RDMs,
 which are obtained in the two cases with and without subsystem-environment interaction, respectively.
 The result shows closeness of the two RDMs and, as a consequence, to the Gibbs state,
 when the interaction is relatively weak.

 Two problems remain open related to the approach of  Ref.~\cite{RGE12}.
 (i) The obtained results are for a generic environment, independent of whether it undergoes
 a complex motion or not.
 One interesting question is whether the upper bound given there may be significantly lowered for
 complex environments such as quantum chaotic systems.
 And, (ii) it is unclear how this approach may give a practically feasible method of finding
 the Hamiltonian that should be used in the Gibbs state, which may take into account some
 effects of the subsystem-environment interaction.

 To the same problem of relatively weak subsystem-environment interaction,
 in the same year of 2012, a different approach was reported in Ref.~\cite{pre12-sta},
 in which elements of  the RDM in the eigenbasis of the central system's Hamiltonian are studied directly.
 In this approach, a more specific situation is considered, in which the subsystem is
 locally coupled to an environment as a many-body quantum chaotic system
 that satisfies the so-called eigenstate thermalization hypothesis (ETH)
 \cite{srednicki1994ETH,Deutch91,srednicki1999ETH,d2016quantum,Deutch18}.
 And, closeness is shown between  the typical-state-computed RDM and the Gibbs state.
 This approach gives an explicit expression for the Hamiltonian that should be
 used in the Gibbs state, which takes into account certain averaged effect of the interaction.

 Two problems remain open related to the approach of Ref.~\cite{pre12-sta}.
 (a) No upper bound was derived explicitly
 for the difference between elements of the studied RDM and those of the Gibbs state.
 For this reason, although this approach and that of Ref.~\cite{RGE12} reach the same qualitative conclusion of
 closeness of the RDM to the Gibbs state under relatively weak interactions,
 a quantitative comparison of their predictions for the condition and extent of the closeness is unavailable.
 And (b) an upper bound for the width of energy eigenfunctions (EFs) of the total system
 in the uncoupled basis was derived and made use of in Ref.~\cite{pre12-sta},
 based on a first-order perturbation-theory treatment to long tails of the EFs.
 Although it was pointed out there that this perturbative treatment may be justified by a
 generalized Brillouin-Wigner perturbation theory \cite{WIC98,pre00-02-GBW},
 a detailed analysis was not given.
 \footnote{ We are to give a further study for this problem in a different paper \cite{EF-semp}.}

 More recently, a relationship was found among elements of the long-time
 averaged RDM of a qubit, which is locally coupled to
 a many-body quantum chaotic system that initially lies in a typical state
 within an energy shell \cite{YWW-decoh}.
 This relationship shows the existence of some nontrivial off-diagonal elements of RDM.
 It is unclear whether the above-discussed two approaches may accommodate this type of
 relationship among elements of RDM.

\subsection{Problems to be studied and organization of the paper}

 In this paper, we are to derive upper bounds related to the first remaining problem of the
 second approach discussed above, but, in a situation more generic than that discussed in Ref.~\cite{pre12-sta}.
 We first discuss MC-ensemble-computed RDMs, then, discuss typical-state-computed RDMs.

 Specifically, we are to consider a generic, isolated, and large quantum system, which is described by an MC ensemble
 (or by a typical state) within an energy shell.
 The isolated (total) system is divided into a generic, small, central subsystem and a large environment,
 with a generic type of interaction.
 We are to derive upper bounds for the difference between elements of the RDM of the
 central subsystem and those elements that are obtained under vanishing interaction.
 The derived expressions are written with properties of the systems involved,
 such as the width of the energy shell, level spacings of the subsystem,
 the maximum width of total EFs, and so on.

 The derived upper bounds are valid in a wide region of the interaction strength, from very weak to strong.
 (i) For very weak interactions, we are to check whether the generic results to be derived are in consistency with
 the known fact that the RDM is close to a Gibbs state \cite{Goldstein06,RGE12,pre12-sta}.
 (ii) For relatively weak interactions, we are to compare the obtained results with results given in Refs.~\cite{RGE12,pre12-sta},
 to see whether the latter are complete.
 In fact, we are to show that the latter are incomplete.
 (iii) For relatively-weak and strong interactions, a sufficient condition will be given,
 under which a considered RDM may have  a Gibbs form.
 As an application, we are to discuss environments as many-body quantum chaotic systems.

 The paper is organized as follows.
 In Sec.~\ref{sect-setup}, we describe the basic framework for our study.
 In Sec.~\ref{sect-diagRDM-MC}, we derive upper bounds for the difference between diagonal elements
 of two MC-ensemble-computed RDMs, which are obtained with and
 without the subsystem-environment interaction, respectively.
 Then, we discuss some applications of the obtained results
 and compare them with a prediction of Ref.~\cite{RGE12}.

 In Sec.~\ref{sect-offdiagRDM-MC},
 we derive an expression for the difference between off-diagonal elements of the above-mentioned RDMs,
 then, as an illustration, we discuss a simple example
 with a two-level system as the central subsystem and a many-body quantum chaotic system as the environment.
 After that, we compare the obtained results with some predictions of Refs.~\cite{RGE12,pre12-sta,YWW-decoh}.
 In Sec.~\ref{sect-reno-Gibbs-S}, as an application,
 we give a sufficient condition under which the RDM of an interacting subsystem may have a canonical Gibbs form,
 when the interaction is not weak; here, particular attention is paid to environments as many-body quantum
 chaotic systems.
 In Sec.~\ref{sect-RDM-typical},
 we discuss differences between elements of typical-state-computed
 and MC-ensemble-computed RDMs.
 Finally, conclusions and discussions are given in Sec.~\ref{sect-conclusion}.

\section{The setup}\label{sect-setup}

 In this section, we give the basic framework, within which we are to give our discussions.
 In Sec.~\ref{sect-H}, we discuss basic properties of the systems to be studied, particularly
 their Hamiltonians and eigenstates.
 In Sec.~\ref{sect-energy-shell}, we discuss basic properties of the MC ensemble for the total system,
 as well as those of the RDM of the central subsystem.

\subsection{Hamiltonians and their eigenstates}\label{sect-H}

 We consider a generic, isolated, and large quantum system, denoted by $\T$, which is divided into
 a small subsystem denoted by $S$ and a large environment denoted by $\E$.
 The Hilbert spaces of $S$ and $\E$ are denoted by $\mathcal{H}^S$ and $\mathcal{H}^\E$,
 respectively, with dimensions $d_S$ and $d_\E$.
 The total Hamiltonian is written as
 \begin{equation}\label{H}
 H = H^{S} + H^{I} + H^{\E},
 \end{equation}
 where $H^S$ and $H^\E$ are the self-Hamiltonians of $S$ and $\E$,
 respectively, and $H^I$ represents the interaction.
 Note that, more precisely, say, $H^S$ on the right-hand side (rhs) of Eq.~(\ref{H}) should be written as
 $H^S \otimes I^\E$, where $I^\E$ represents the identity operator acting on $\mathcal{H}^\E$;
 but, for brevity, we usually omit the identity operator.
 We use $H^0$ to denote the uncoupled Hamiltonian of the total system, i.e.,
\begin{gather}\label{H0}
 H^0 = H^S + H^\E.
\end{gather}
 It is sometimes convenient to introduce a parameter for characterizing the strength of the interaction;
 in this case, we use the Greek letter  $\lambda$, with $ \| H^I \| \propto \lambda$.

 The interaction Hamiltonian $H^I$ is of a generic type for the main results to be given in
 Secs.~\ref{sect-diagRDM-MC} and \ref{sect-offdiagRDM-MC}.
 In some applications of these results, which will be given in Secs.~\ref{sect-TLS} and \ref{sect-reno-Gibbs-S},
 local interactions are considered as explicitly indicated there.
 In the derivation of the main results, there is only one requirement for the $S$-$\E$ interaction strength,
 which is that $\rho_{\rm dos}^H \simeq \rho_{\rm dos}^{H^0}$,
 where $\rho_{\rm dos}^H$ and $\rho_{\rm dos}^{H^0}$ represent the density of states of the total
 system with Hamiltonians $H$ and $H^0$, respectively.
 This requirement, which implies that the influence of the $S$-$\E$ interaction in the density of states of
 the total system can be neglected, is satisfied in almost all situations of practical interest
 with large environments.

 Normalized eigenstates of the total Hamiltonian $H$ are denoted by $|n\ra$ with energies $E_n$
 in the increasing-energy order,
\begin{gather}\label{H|n>}
 H|n\rangle = E_n|n\rangle.
\end{gather}
 Normalized eigenstates of $H^S$ are denoted by $|\alpha\rangle$, with energies $e^S_\alpha$,
 and those of $H^\E$ by $|i\rangle$ with energies $e_i$,
 both in the increasing-energy order,
\begin{eqnarray} \label{Sa}
H^S|\alpha\rangle &=& e^S_\alpha|\alpha\rangle, \\
H^\E|i\rangle &=& e_i|i\rangle, \label{Ei}
\end{eqnarray}
 where for brevity we have omitted a superscript $\E$ for the environmental energy $e_i$.

 The RDM to be studied is written on the eigenbasis of $H^S$.
 When the spectrum of $H^S$ has some degeneracy, an ambiguity exists in writing the related eigenstates of $H^S$;
 to fix this ambiguity, some additional requirement is needed.
 To avoid this complexity, we assume that the system $S$ has a nondegenerate spectrum.
 \footnote{In fact, discussions to be given in Sec.~\ref{sect-diagRDM-MC} for diagonal elements are independent of whether
 the spectrum of $S$ is degenerate or not,
 while the main results of Sec.~\ref{sect-offdiagRDM-MC} for off-diagonal elements 
 are invalid for $H^S$ with a degenerate spectrum
 [see, e.g., Eq.~(\ref{rhoS-rhoSn-ov})].}

 Eigenstates of $H^0$ with eigenenergies $E_{\alpha i} $ are written as
 $|\alpha\rangle |i\rangle$, in short $|\alpha i \rangle$,  satisfying
\begin{gather}\label{}
 H^0|\alpha i\ra = E_{\alpha i}|\alpha i\ra, \quad E_{\alpha i} = e^S_\alpha + e_i.
\end{gather}
 In the energy order, the states $|\alpha i\ra$ are indicated by $|E_r\ra$ with one integer label $r$,
 which has a one-to-one correspondence to the pair $(\alpha, i)$, namely, $r \leftrightarrow (\alpha, i)$,
 such that $E_r = E_{\alpha i}$ and
\begin{gather}\label{}
 H^0|E_r\ra = E_{r}|E_r\ra, \quad  E_r \le E_{r+1}.
\end{gather}
 Expansions of the states $|n\ra$ in the bases $|\alpha i\ra$ and $|E_r\ra$, with coefficients
 denoted by $C_{\alpha i}^n$ and $C_r^n$, respectively, are written as
 \begin{equation}\label{|n>}
 |n\rangle = \sum_{\alpha, i}C_{\alpha i}^n|\alpha i\rangle = \sum_{r}C_{r}^n|E_r\rangle.
\end{equation}
 The coefficients $C_{\alpha i}^n$ and $C_r^n$ give the EFs.

 Significant components $C_r^n$ of a given state $|n\ra$ usually occupy a restricted region in the uncoupled
 spectrum, say, in a region of $ E_r$ with $r$ between $r^{(n)}_1$ and $r^{(n)}_2$.
 For brevity, we call such a region \emph{a ``main-body'' region} of $|n\ra$.
 To characterize a main-body region, one may employ a small positive parameter $\epsilon$,
 such that the population of $|n\ra$ outside this region is smaller than $\epsilon$.
 \footnote{The exact value of $\epsilon$ is usually case-dependent.
 That is, it depends on what is needed for the problem at hand; it may be, say, $10\%$, or $1\%$.
 But, the name of ``main body'' implies that one should not take $\epsilon = 0$. }
 We use $\Omega_n$, $\Omega_n  \equiv [r^{(n)}_1, r^{(n)}_2]$, to indicate such a region, for which
\begin{gather}\label{main-body}
 \sum_{r \in \Omega_n} |\la E_r|n\ra |^2 \doteq 1-\epsilon,
\end{gather}
 where ``$\doteq$'' means that the left-hand side is either equal to the rhs, or is just larger than  the rhs,
 such that it become smaller than the rhs when $\Omega_n$ is shrunk
 by letting $r^{(n)}_1 \to r^{(n)}_1 +1$ or $r^{(n)}_2 \to r^{(n)}_2 -1$.
 We use $w_E$ to denote the maximum width of the energy region occupied by $\Omega_n$, i.e.,
\begin{gather}\label{w-E}
 w_E= \max \left\{ \left( E_{r^{(n)}_2} - E_{r^{(n)}_1} \right)\right\}
\end{gather}
 for those states $|n\ra$ that lie in the energy region of the total system of relevance to our discussions to be given later.

 The so-called local spectral density of states (LDOS), or strength function in nuclear physics,
 will also be used in our later discussions.
 They are the reverse of EFs, that is,
 the LDOS of an uncoupled state $|E_r\ra$ is given by its expansion in the basis of $\{|n\ra \}$.
 We use $\Omega_r^L$ to denote a main-body region of $|E_r\ra$,
 which is written as $\Omega_r^L \equiv [n^{(r)}_1, n^{(r)}_2]$ for a region of the label
 $n$ between $n^{(r)}_1$ and $n^{(r)}_2$; it satisfies the following relation:
\begin{gather}\label{main-body-L}
 \sum_{n \in \Omega_r^L} |\la E_r|n\ra |^2 \doteq 1-\epsilon.
\end{gather}
 The maximum value of the energy width of $\Omega_r^L$, namely, of $(E_{n^{(r)}_2} - E_{n^{(r)}_1})$,
 for those states $|E_r\ra$ in the energy region of relevance, is denoted by $w_L$.
 We use $w_M$ to indicate the larger one of $w_E$ and $w_L$, namely,
\begin{gather}\label{}
 w_M = \max \{ w_E, w_L \}.
\end{gather}

 For a sufficiently small $\epsilon$,
 the value of $E_n$ lies within  the main-body energy region of the EF of $|n\ra$,
 meanwhile, $E_r$ lies within the main-body region of the LDOS of $|E_r\ra$.
 \footnote{We neglect the trivial case of $[H^S+H^\E, H^I]=0$, in which the
 states $|n\ra$ are equal to the uncoupled ones $|E_r\ra$.  }
 It is not difficult to verify that these two properties imply the following relations, respectively,
\begin{subequations}\label{EF-LOS-region}
\begin{gather}
 E_n -w_M \le E_{r^{(n)}_2} < E_{r^{(n)}_1} \le E_n + w_M, \label{EF-region}
 \\ E_r -w_M \le E_{n^{(r)}_2} < E_{n^{(r)}_1} \le E_r + w_M. \label{LDOS-region}
\end{gather}
\end{subequations}
 That is, the main body of the EF of $|n\ra$ lies within the region of $E_r \in [E_n -w_M, E_n + w_M]$,
 meanwhile, the main body of the LDOS of $|E_r\ra$ lies within the region of $E_n \in [E_r -w_M, E_r + w_M]$.

\subsection{MC Energy shell and RDM}\label{sect-energy-shell}

 We consider an MC-ensemble description of the total system
 within an energy shell denoted by $\Gamma$, which starts at an energy denoted by $E_s$
 and has a width $\Delta$, i.e., $\Gamma  = [E_s, E_s+\Delta ]$,
 with the subscript ``s'' standing for ``starting of shell''.
 The energy shell $\Gamma$ is far from edges of the spectrum of the total system.
 We use $\HH_{\Gamma }$ to denote the subspace spanned by those eigenstates $|n\ra$
 with $E_n \in \Gamma $.
 The dimension of $\HH_{\Gamma }$ is denoted by $d_{\Gamma}$.
 In statistical physics, the energy shell $\Gamma$, though narrow, is assumed to be wide enough to contain
 very many levels $E_n$.

 The MC description of the total system within the energy shell $\Gamma$ is written as
\begin{gather}\label{rho-T}
 \rho^\T =  \frac{1}{d_{\Gamma}} \sum_{E_n \in \Gamma } |n\ra \la n|.
\end{gather}
 The RDM of the system $S$, denoted by $\rho^S$, is given by
 \begin{equation}\label{}
 \rho^S \equiv \tr_\E \left (  \rho^\T \right ) .
 \end{equation}
 Its elements are written as
\begin{gather}\label{rhoS-rhoSn}
 \rho^S_{\alpha \beta} \equiv \la \alpha |\rho^S|\beta\ra
 = d_{\Gamma }^{-1} \sum_{E_n \in \Gamma} \rho^{S(n)}_{\alpha \beta},
\end{gather}
 where $\rho^{S(n)}_{\alpha \beta}$ indicate elements computed from a
 single eigenstate $|n\ra$, i.e.,
\begin{gather}\label{rhoSn}
  \rho^{S(n)}_{\alpha \beta} \equiv  \la \alpha|\tr_\E (|n\ra \la n|)|\beta\ra
   = \sum_i   C^{n}_{\alpha i} C^{n*}_{\beta i}.
\end{gather}

 For the uncoupled system $H^0$, one may consider a similar energy shell
 denoted by $\Gamma^0$, with $\Gamma^0  = [E_s, E_s+\Delta ]$.
 We use $d_{\Gamma^0}$ to denote the number of levels $E_r$ within $\Gamma^0$.
 The MC ensemble in the uncoupled case is described by
\begin{gather}\label{}
 \rho^{\T 0} =  \frac{1}{d_{\Gamma^0}} \sum_{E_r \in \Gamma^0 } |E_r\ra \la E_r|.
\end{gather}
 This gives the RDM $ \rho^{S0} \equiv \tr_\E \left (  \rho^{\T0} \right )$, with elements
 $\rho^{S0}_{\alpha \beta} \equiv \la \alpha |\rho^{S0}|\beta\ra $.
 For a given state $|\alpha\ra$ of the system $S$, we use $\Gamma^{\E}_{\alpha}$ to denote
 the environmental energy shell, which contains those environmental levels $e_i$ for which
 $ E_{\alpha i} =E_r \in \Gamma^0$, i.e.,
\begin{gather}\label{Gamma-Ea}
 \Gamma^{\E}_{\alpha}  = [E_s - e^S_\alpha, E_s- e^S_\alpha+\Delta ].
\end{gather}
 We use $\HH^\E_{\Gamma \alpha}$ to denote the subspace spanned by $|i\ra \in \Gamma_{ \alpha}^\E$
 and use $d^{\E}_{\Gamma\alpha}$ to indicate its dimension.

 It is straightforward to find that
\begin{subequations}\label{rhoS0-solu}
\begin{gather}\label{rhoS0-aa}
 \rho^{S0}_{\alpha \alpha}= \frac{1}{d_{\Gamma^0}}  d^{\E}_{\Gamma\alpha}, \quad \forall \alpha,
 \\ \rho^{S0}_{\alpha \beta}=0, \qquad \quad \forall \alpha \ne \beta.
\end{gather}
\end{subequations}
 Then, under the well-known assumption about an exponential shape of the density of states,
 one gets that
\begin{subequations}\label{rhoS0-rhoG}
\begin{gather}\label{}
 \rho^{S0}_{\alpha \alpha} \simeq {( \rho^S_{G} )}_{\alpha \alpha} \qquad \quad \forall \alpha,
 \\ \rho^{S0}_{\alpha \beta}={(\rho^S_{G})}_{\alpha \beta}=0 \qquad \quad \forall \alpha \ne \beta,
\end{gather}
\end{subequations}
 where $\rho^S_{G}$ indicates the Gibbs state,
\begin{gather}\label{Gibbs}
  \rho^S_{G}  = e^{-\beta H^S}/ \tr e^{-\beta H^S},
\end{gather}
 with a parameter $\beta$ determined by the density of states of the environment.
 Hence, instead of studying the differences $|\rho^{S}_{\alpha \beta} - (\rho^S_{G})_{\alpha \beta}|$,
 below we study $|\rho^{S}_{\alpha \beta} - \rho^{S0}_{\alpha \beta}|$.

\section{Difference between diagonal elements of RDMs} \label{sect-diagRDM-MC}

 In this section, we discuss diagonal elements of RDMs, under generic $S$-$\E$ interactions
 with only one restriction, i.e., $\rho_{\rm dos}^H \simeq \rho_{\rm dos}^{H^0}$. 
 In Sec.~\ref{sect-MC-DRM-diagonal}, we derive upper bounds  for
 $|\rho^S_{\alpha \alpha} - \rho^{S0}_{\alpha \alpha}|$ in the case of $\Delta >2 w_M$.
 Since an MC energy shell $\Gamma$ should contain very many levels,
 it is this case of $\Delta > 2 w_M$ that is often met in statistical physics.
 \footnote{Besides properties of the eigenstates of the systems involved, 
 the width $w_M$ is also determined by the parameter $\epsilon$.
 Practically, $\epsilon$ does not need to take a very small value.}
 The opposite case of $\Delta < 2 w_M$ is discussed in Sec.~\ref{sect-MC-DRM-diagonal-ww}.
 Finally, in Sec.~\ref{sect-diag-appl}, we discuss some applications of the results obtained.

\subsection{Upper bounds of $|\rho^S_{\alpha \alpha} - \rho^{S0}_{\alpha \alpha}|$ for $\Delta >2 w_M$}
\label{sect-MC-DRM-diagonal}

 In this section, for $\Delta > 2 w_M$,
 we derive the following expression for $(\rho^S_{\alpha \alpha} - \rho^{S0}_{\alpha \alpha})$,
 in the case that linear approximation is valid for 
 the environmental density of states around the energy shell $\Gamma^{\E}_{\alpha} $.
 The expression is
\begin{gather}\label{drho-simeq}
 \rho^{S}_{\alpha \alpha} - \rho^{S0}_{\alpha \alpha} \simeq
 q_1 \frac{ w_M }{ \Delta} \frac{d^{\E}_{\Gamma\alpha}}{d_\Gamma} + q_0\epsilon,
\end{gather}
 where $q_1$ and $q_0$ are two undetermined parameters satisfying $|q_1| <2$ and $|q_0|<1$. 
 (See Eq.~(\ref{q1-q0}) to be given below for explicit expressions of $q_1$ and $q_0$.)
 The opposite case with invalidity of the linear approximation, which is not often met for
 narrow energy shells,  is briefly addressed at the end of this section.

 Making use of Eq.~(\ref{drho-simeq}) and noting that $d^{\E}_{\Gamma\alpha} < d_\Gamma$,
 one gets the following upper bound for the diagonal difference, 
\begin{gather}\label{drho-aa-1}
 |\rho^{S}_{\alpha \alpha} - \rho^{S0}_{\alpha \alpha}|  \lesssim  \frac{2 }{ \Delta}w_M + \epsilon.
\end{gather}
 When the level spacings of the system $S$ are small, 
 the differences among $d^{\E}_{\Gamma\alpha}$ of different $\alpha$ may be
 small compared with the values of $d^{\E}_{\Gamma\alpha}$;
 in this case, one has $d_\Gamma \simeq d_S d^{\E}_{\Gamma\alpha}$ and
 an estimate better than Eq.~(\ref{drho-aa-1}) can be obtained, i.e., 
\begin{gather}\label{drho-aa}
 |\rho^{S}_{\alpha \alpha} - \rho^{S0}_{\alpha \alpha}|  \lesssim  \frac{2 }{d_S \Delta}w_M + \epsilon.
\end{gather}

 Below, we give the derivation for Eq.~(\ref{drho-simeq}),
 which is valid in the case of the above-mentioned linear approximation. 
 To this end, 
 we divide the environmental spectrum $\{ e_i \}$  into several regions separated by the following parameters,
\begin{subequations}\label{varepsilon}
\begin{gather}
 \varepsilon_1 = E_s-e^S_\alpha-w_M ,
 \\ \varepsilon_2 =  \varepsilon_1  + 2w_M ,  \label{varepsilon-2}
 \\ \varepsilon_3 =  \varepsilon_1  + \Delta  ,  \label{varepsilon-3}
 \\ \varepsilon_4 =  \varepsilon_2  + \Delta .
\end{gather}
\end{subequations}
 We use $\R_{\kappa }^{\E\alpha}$ with $\kappa =0,1,2,3$ to denote the following four regions of the spectrum
 separated by the above parameters, i.e.,
\begin{subequations}\label{R-Ea-k}
\begin{gather}
 \R_{0}^{\E\alpha} := [e_{\rm start}, \varepsilon_1)  \cup (\varepsilon_4,e_{\rm end}],
 \\ \R_{1}^{\E\alpha} := [\varepsilon_1 , \varepsilon_2),
 \\ \R_{2}^{\E\alpha} := [\varepsilon_2 ,  \varepsilon_3 ), \label{R-Ea-2}
 \\  \R_{3}^{\E\alpha} := [ \varepsilon_3 ,  \varepsilon_4],
\end{gather}
\end{subequations}
 where $e_{\rm start}$ and $e_{\rm end}$ indicate the starting and ending levels of the environmental spectrum, respectively.
 It is seen that the region $\R_{2}^{\E\alpha}$ lies inside the energy shell $\Gamma^{\E}_{\alpha}$,
 with a width $(\Delta - 2w_M)$;
 the two regions of $\R_{1}^{\E\alpha}$ and $\R_{3}^{\E\alpha}$ lie at the two borders of the shell, respectively, 
 each with a width $2w_M$;
 and the region $\R_{0}^{\E\alpha}$ lies completely outside the shell.

 With the above-discussed division of the environmental spectrum, 
 making use of Eqs.~(\ref{rhoS-rhoSn}) and (\ref{rhoSn}), the diagonal element
 $\rho^{S}_{\alpha \alpha} $ is written as
\begin{gather}\label{K1a-2}
  \rho^{S}_{\alpha \alpha} = d_\Gamma^{-1} \sum_{\kappa =0}^{3} F_{\alpha \kappa  },
\end{gather}
 where
\begin{gather}\label{Gak}
  F_{\alpha \kappa  } = \sum_{e_i\in \R_{\kappa }^{\E\alpha}} \sum_{E_n \in \Gamma}   |C^n_{\alpha i}|^2.
\end{gather}
 We use $N^{\E\alpha}_{\kappa }$ to denote the number of those levels $e_i$ that
 lie within a region $\R_{\kappa }^{\E\alpha}$.

 We discuss contributions from the four regions $\R_{\kappa }^{\E\alpha}$ separately.
 Firstly, we discuss the central region $\R_{2}^{\E\alpha}$, which usually gives the main contribution
 to $\rho^{S}_{\alpha \alpha} $.
 We write $F_{\alpha 2}$ in the following form,
\begin{gather}\label{Fa2-1}
 F_{\alpha 2} = N_{2}^{\E\alpha} + \sum_{e_i\in \Gamma_{\alpha 2}^\E} (I_{\alpha i} -1),
\end{gather}
 where
\begin{gather}
 I_{\alpha i} = \sum_{E_n \in \Gamma}   |C^n_{\alpha i}|^2. \label{Iai}
\end{gather}
 For a level $e_i \in \R_{2}^{\E\alpha}$, according to Eqs.~(\ref{varepsilon-2}), (\ref{varepsilon-3}),
 and (\ref{R-Ea-2}),  the value of $E_r = E_{\alpha i}$
 lies between $(E_s + w_M)$ and $(E_s  + \Delta  - w_M)$.
 Due to Eq.~(\ref{LDOS-region}), this implies that
 the main-body region of  the LDOS of $|\alpha i\ra$ should lie within the energy shell $\Gamma$.
 Hence, $(1-I_{\alpha i}) \le \epsilon$ [see Eq.~(\ref{main-body-L})].
 As a result, $F_{\alpha 2} $ in Eq.~(\ref{Fa2-1}) can be written as
\begin{gather}\label{Fa-2}
 F_{\alpha 2} = N_{2}^{\E\alpha} - a_2 \epsilon N_{2}^{\E\alpha},
\end{gather}
 where $a_2$ is some undetermined real parameter satisfying $0< a_2 <1$ .

 Next, we discuss the two regions $\R_{\kappa }^{\E\alpha}$ of $\kappa =1$ and $3$,
 each with a width $2w_M$.
 For some of the levels $e_i$ lying within these two regions, the values of $I_{\alpha i}$ are close to $1$,
 meanwhile,  for some other levels $I_{\alpha i}$ are much smaller than $1$.
 Since the environmental density of states around the energy shell $\Gamma^{\E}_{\alpha} $
 is approximately a linear function, its average value is approximately given by $({d^{\E}_{\Gamma\alpha}}/{\Delta})$.
 Then,  $F_{\alpha \kappa}$ of $\kappa =1,3$ can be written in the following form,
\begin{gather}\label{Fa-13}
 F_{\alpha \kappa} = 2a_{\kappa}   w_M d^{\E}_{\Gamma\alpha} /\Delta , \quad \kappa =1,3,
\end{gather}
 where $a_1$ and $a_3$ are some undetermined parameters satisfying $0< a_{1(3)} <1$.
 In most cases, the values of $a_1$ and $a_3$ are around $0.5$ or smaller.

 Finally, we discuss the region $\R_{0}^{\E\alpha}$.
 For an energy level $e_i$ lying in this region, the value of $E_r = e_i + e^S_\alpha$ is either
 smaller than $(E_s-w_M)$, or larger than $(E_s + \Delta +w_M)$.
 This implies that $E_r$ lies outside the main-body regions of all those states  $|n\ra \in \Gamma$.
 Hence, according to Eq.~(\ref{main-body}), one has
\begin{gather}\label{cna2<ep}
 \sum_{e_i\in \R_{0}^{\E\alpha}}  |C^n_{\alpha i}|^2  < \epsilon.
\end{gather}
 This gives the following expression:
\begin{gather}\label{Fa-0}
 F_{\alpha 0 } = \sum_{E_n \in \Gamma} \sum_{e_i\in \R_{0}^{\E\alpha}}  |C^n_{\alpha i}|^2
 = a_0 \epsilon d_\Gamma ,
\end{gather}
 with some undetermined parameter $a_0$ satisfying $0< a_{0} <1$.

 Substituting the above-obtained results for $F_{\alpha \kappa }$ into Eq.~(\ref{K1a-2}), one gets that
\begin{gather}\label{rhos-aa-compl}
  \rho^{S}_{\alpha \alpha} = \frac{N_{2}^{\E\alpha}}{d_\Gamma}
  +  \frac{2(a_1+a_3) w_M}{\Delta}  \frac{d^{\E}_{\Gamma\alpha}}{d_\Gamma}
  + (a_0  -  a_2 \frac{N_{2}^{\E\alpha}}{d_\Gamma}) \epsilon.
\end{gather}
 To go further, we make use of the assumption of $\rho_{\rm dos}^H \simeq \rho_{\rm dos}^{H^0}$,
 i.e., the difference between the density of states of $H^0$
 and that of $H$ can be neglected; this implies that $d_{\Gamma^0} \simeq d_{\Gamma}$.
 Moreover, we note that $(d^{\E}_{\Gamma\alpha}- N_{2}^{\E\alpha})$ 
 is equal to the number of levels that lie inside the overlap of the energy shell $\Gamma^\E_\alpha$ 
 and the two regions of $\R_{1}^{\E\alpha}$ and $\R_{3}^{\E\alpha}$.
 Then, due to the validity of linear approximation for the environmental density of states 
 within the energy shell $ \Gamma^{\E}_{\alpha}$, it is easy to see that
 $d^{\E}_{\Gamma\alpha}- N_{2}^{\E\alpha} \simeq 2 w_M (d^{\E}_{\Gamma\alpha}/\Delta )$.
 Making use of these properties, from Eqs.~(\ref{rhos-aa-compl}) and (\ref{rhoS0-aa}), one finds that
\begin{gather}\label{}\notag
 \rho^{S}_{\alpha \alpha} - \rho^{S0}_{\alpha \alpha}
  \simeq  \frac{2 (a_1+a_3) w_M}{\Delta}  \frac{d^{\E}_{\Gamma\alpha}}{d_\Gamma}
 -\frac{2 w_M d^{\E}_{\Gamma\alpha}}{d_\Gamma \Delta}
  \\ +  (a_0  -  a_2 \frac{N_{2}^{\E\alpha}}{d_\Gamma}) \epsilon .
\end{gather}
 This finishes the derivation of Eq.~(\ref{drho-simeq}), with the following relations,
\begin{gather}\label{q1-q0}
 q_1 = 2 (a_1+a_3) -2, \quad q_0 = a_0  -  a_2 \frac{N_{2}^{\E\alpha}}{d_\Gamma}.
\end{gather}

 One remark: The upper bounds  for $| \rho^{S}_{\alpha \alpha} - \rho^{S0}_{\alpha \alpha}|$ 
 given in Eqs.~(\ref{drho-aa-1}) and (\ref{drho-aa}) correspond to the maximum value of $|q_1|$.
 For a concrete system in which the value of $q_1$ can be evaluated, 
 Eq.~(\ref{drho-simeq}) may give a much lower upper bound.
 For example, as mentioned above, $a_1$ and $a_3$ may be around $0.5$ in some systems;
 in such a system, $q_1$ is small and this may considerably reduce
 $| \rho^{S}_{\alpha \alpha} - \rho^{S0}_{\alpha \alpha}|$ according to Eq.~(\ref{drho-simeq}).

 Finally, we give a brief discussion for the case that linear approximation is invalid for 
 the environmental density of states within the energy shells $\Gamma^{\E}_{\alpha} $. 
 In this case, Eqs.~(\ref{Fa-2}) and (\ref{Fa-0}) are still valid, while, Eq.~(\ref{Fa-13}) is replaced by
\begin{gather}\label{Fa-13-se}
 F_{\alpha \kappa} = 2a_{\kappa}   w_M  \rho^\E_{\rm dos, \kappa}, \quad \kappa =1,3,
\end{gather}
 where $\rho^\E_{\rm dos, \kappa}$ indicates the environmental density of states 
 in the region $\R_{\kappa }^{\E\alpha}$.
 Then, after simple derivations,  one gets the following estimate,
\begin{gather}\label{rhod-nonlinear}
 |\rho^{S}_{\alpha \alpha} - \rho^{S0}_{\alpha \alpha}| \lesssim 
 \frac{(\rho^\E_{\rm dos, 1} + \rho^\E_{\rm dos, 3} )}{d_\Gamma}w_M + \epsilon.
\end{gather}

 A final remark: The upper bounds given above in Eqs.~(\ref{drho-aa-1}), (\ref{drho-aa}), and (\ref{rhod-nonlinear})
 show the same dependence on $\epsilon$ and $w_M$, with differences only in the prefactors of $w_M$. 
 
\subsection{Upper bounds of $|\rho^S_{\alpha \alpha} - \rho^{S0}_{\alpha \alpha}|$ for $\Delta < 2 w_M$}
\label{sect-MC-DRM-diagonal-ww}

 In this section, we discuss $|\rho^S_{\alpha \alpha} - \rho^{S0}_{\alpha \alpha}|$ for $\Delta < 2 w_M$.
 In the case that  linear approximation is valid for 
 the environmental density of states around the energy shell $\Gamma^{\E}_{\alpha} $, we are to show that
\begin{gather}\label{drho-aa-2-ww}
 |\rho^{S}_{\alpha \alpha} - \rho^{S0}_{\alpha \alpha}| \lesssim 
 \frac{2 w_M}{ \Delta}  \frac{d^{\E}_{\Gamma\alpha}}{d_\Gamma}    + \epsilon.
\end{gather}
 Then, it is easy to see that the upper bounds given in Eqs.~(\ref{drho-aa-1}) and (\ref{drho-aa}) for 
 $|\rho^{S}_{\alpha \alpha} - \rho^{S0}_{\alpha \alpha}|$
 are, in fact, valid independent of the relation between $\Delta$ and $2 w_M$.
 The opposite case with invalidity of the linear approximation is briefly discussed at the end of this section.

 To deal with the case with validity of the linear approximation,
 basically, one may follow a procedure similar to that adopted in the previous section.
 Note that, with $\Delta < 2 w_M$, 
 the previously discussed subregion $\R_{2}^{\E\alpha}$ should shrink to zero.
 Hence, when dividing the environmental spectrum into subregions $\R_{\kappa }^{\E\alpha}$,
 we use the following values of the parameters $\varepsilon_\kappa$,
\begin{subequations}\label{varepsilon-ww}
\begin{gather}
 \varepsilon_1 = E_s-e^S_\alpha-w_M ,
 \\ \varepsilon_2 = \varepsilon_3 =  E_s-e^S_\alpha + \frac 12 \Delta ,  \label{varepsilon-2-ww}
 \\ \varepsilon_4 =   E_s-e^S_\alpha   + \Delta + w_M.
\end{gather}
\end{subequations}
 It is seen that the positions of $\varepsilon_1$ and $\varepsilon_4$ are unchanged,
 and $\varepsilon_2= \varepsilon_3$ indicate the middle of the energy window $\Gamma^\E_\alpha$;
 as a result, $\R_{0}^{\E\alpha}$ remains unchanged and $\R_{2}^{\E\alpha}$ is empty.

 The elements $\rho^{S}_{\alpha \alpha}$ are also written as in Eq.~(\ref{K1a-2})
 and can be studied by the same method as that used previously.
 It is easy to see that $F_{\alpha 2} =0$ and $F_{\alpha 0} $ remains unchanged.
 For $F_{\alpha \kappa} $ of $\kappa =1,3$, 
 when the linear approximation is valid for the environmental density of states,  similar to Eq.~(\ref{Fa-13}), we find that
\begin{gather}\label{F13-2}
 F_{\alpha \kappa} = a_{\kappa}  (w_M+ \Delta /2) \frac{d^{\E}_{\Gamma\alpha}}{\Delta} , \quad \kappa =1,3,
\end{gather}
 where the parameters $a_\kappa$ have properties similar to those discussed previously.
 Putting these $F_{\alpha \kappa} $ together, one gets that
\begin{gather}\label{rhos-aa-ww}
  \rho^{S}_{\alpha \alpha} =   (a_1+a_3) \left( \frac{w_M}{\Delta}+ \frac 12 \right) \frac{d^{\E}_{\Gamma\alpha}}{d_\Gamma}
  + a_0  \epsilon.
\end{gather}
 This gives that
\begin{gather}\label{rhos-aa-ww2}
  \rho^{S}_{\alpha \alpha} -  \rho^{S0}_{\alpha \alpha}
  \simeq \left(   \frac{ (a_1+a_3)w_M }{ \Delta}
  + \frac{a_1+a_3}2 -1 \right ) \frac{d^{\E}_{\Gamma\alpha}}{d_\Gamma}  + a_0  \epsilon.
\end{gather}
 Due to the facts that $\Delta < 2 w_M$ and $0< a_{1,3}<1$, 
 an upper bound for the absolute value of the term within the big
 parentheses on the rhs of Eq.~(\ref{rhos-aa-ww2}) is obtained with $a_1 = a_3=1$. 
 This gives the estimate in Eq.~(\ref{drho-aa-2-ww}).

 Finally, we briefly discuss the case that linear approximation is invalid for 
 the environmental density of states around the energy shell $\Gamma^{\E}_{\alpha} $.
 In this case, $F_{\alpha 2} $ and $F_{\alpha 0} $ are the same as those discussed above,
 while, $F_{\alpha \kappa}$ of $\kappa =1,3$ in Eq.~(\ref{F13-2}) is replaced by 
 $F_{\alpha \kappa} = a_{\kappa}  (w_M+ \Delta /2)
 \rho^\E_{\rm dos, \kappa}$.
 It is seen that $\rho^{S}_{\alpha \alpha}$ and $\rho^{S0}_{\alpha \alpha}$ share no common item,
 as a result, no concise expression is found for upper bound of the 
 difference $|\rho^{S}_{\alpha \alpha} - \rho^{S0}_{\alpha \alpha}|$.

\subsection{Some applications of Eq.~(\ref{drho-aa}) }\label{sect-diag-appl}

 In this section, we discuss some applications of Eq.~(\ref{drho-aa}).
 We first discuss some main features of its predictions for three regimes of the interaction strength,
 in comparison with results of Refs.~\cite{Goldstein06,RGE12,pre12-sta}.
 Then, we discuss a specific situation, in which EFs have the so-called Breit-Wigner shape.

\subsubsection{Three regimes of interaction strength}

 The estimate in Eq.~(\ref{drho-aa}) was derived under the following conditions:
 (i) $\rho_{\rm dos}^H \simeq \rho_{\rm dos}^{H^0}$,
 (ii) validity of linear approximation to
 the environmental density of states around the energy shell $\Gamma^{\E}_{\alpha} $,
 and (iii) $d_\Gamma \simeq d_S d^{\E}_{\Gamma\alpha}$,
 independent of the relation between $\Delta$ and $2 w_M$.
 Hence, it holds in a wide regime of the $S$-$\E$ interaction strength, from extremely weak to strong.
 Below, we discuss the three regimes of very weak, relatively weak,
 and strong separately.

 (i) Under interactions that are very weak such that $(w_M/d_S \Delta )$ is close to zero, 
 Eq.~(\ref{drho-aa}) predicts that the difference between $\rho^{S}_{\alpha \alpha}$ and $\rho^{S0}_{\alpha \alpha}$
 can be neglected.
 Then, for an environment whose density of states has an exponential shape,
 the diagonal elements of the RDM, namely $\rho^{S}_{\alpha \alpha}$, are quite close to
 those of the Gibbs state $\rho^S_{G}$ in Eq.~(\ref{Gibbs}).
 This is in agreement with the known fact discussed in Refs.~\cite{Goldstein06,RGE12,pre12-sta}.

 (ii) Under relatively weak interactions, for which the ratio $(w_M/ d_S \Delta)$ is not close to zero, but still small, 
 Eq.~(\ref{drho-aa}) can be regarded as a quantitative expression for some qualitative arguments
 used in Ref.~\cite{pre12-sta} to derive main results given there.

 In this relatively weak interaction regime,  the following upper bound is given in Ref.~\cite{RGE12}
 for the trace distance  between the two RDMs $\rho^S$ and $\rho^{S0}$,
 denoted by $\D ( \rho^S , \rho^{S0})$, which appeared as Eq.~(2) there,
\begin{gather}\label{D-REG}
 \D ( \rho^S , \rho^{S0}) \le 4\sqrt{ \frac{\| H^I \|_\infty}{ \Delta }}, \tag{(2)-\cite{RGE12}}
\end{gather}
 where $\| H^I \|_\infty$ indicates the maximum singular eigenvalue of $H^I$ \cite{GE16-thermal-review}.
 The two upper bounds given in Eq.~(\ref{drho-aa}) and Eq.~(\ref{D-REG}), although not identical,
 are qualitatively consistent due to the fact that both $w_M$ and $\| H^I \|_\infty$ are small
 for weak interactions.

 To compare the above-discussed two bounds in a quantitative way, as an example,
 one may consider a special case in which the width $w_M$ is proportional to the parameter
  $\lambda$ for the interaction strength.
 (Another example will be given in the next subsection.)
 If the  $\epsilon$-dependence of $w_M$ is $w_M \propto 1/\epsilon$
 [cf.~Eq.~(\ref{wE-BW}) to be given below],
 then, $w_M \propto (\lambda /\epsilon)$.
 As a result, the rhs of Eq.~(\ref{drho-aa}) has a minimum value proportional to $\sqrt{\lambda / \Delta}$,
 at an appropriate value of the parameter $\epsilon$.
 Then, since $\| H^I \|_\infty \propto {\lambda}$,
 these two upper bounds show the same dependence of $\sqrt{\lambda / \Delta}$.

 (iii) Under interactions that are strong enough for the ratio $(w_M/d_S \Delta)$ to be not small,
 it is possible for  $\rho^{S}_{\alpha \alpha}$ to deviate notably from $\rho^{S0}_{\alpha \alpha}$
 and, as a result, for $\rho^S$ to deviate notably from the Gibbs state $\rho^S_{G}$.
 Even in this case, it is possible for diagonal elements of $\rho^S$ to be close to those of some
 renormalized Gibbs state, which will be discussed in detail in Sec.~\ref{sect-RGS-diag}.

\subsubsection{EFs with a Breit-Wigner form}

 Clearly, the width $w_M$ plays an important role in applications of Eq.~(\ref{drho-aa}).
 Below, as an illustration, we discuss a case that is often met in realistic models,
 in which the EFs and LDOS have on average a Breit-Wigner form \cite{FI00,BM-book},
 described by the following Lorentz function $f(E)$,
\begin{gather}\label{}
 f(E) = \frac{1}{2\pi} \frac{\omega_{BW}}{E^2 + (\omega_{BW}/2)^2},
\end{gather}
 with a width $\omega_{BW}$ given by
\begin{gather}\label{omega-BW}
 \omega_{BW} \simeq 2\pi \ov{| H^I_{rr'}|^2} \rho_{\rm dos},
\end{gather}
 where $\rho_{\rm dos}$ indicates the density of states.
 Under this Breit-Wigner form, making use of Eq.~(\ref{main-body}),
 it is straightforward to find the following expression of $w_E$ and $w_L$,
\begin{gather}\label{wE-BW}
 w_{E,L} = \frac{2\omega_{BW}}{\pi \epsilon}.
\end{gather}
 Clearly, Eq.~(\ref{omega-BW}) predicts that $\omega_{BW} \propto \lambda^2$.

 At a first sight, it seems that the smallness of the parameter $\epsilon$ may imply largeness of $w_E$
  in Eq.~(\ref{wE-BW}).
 However, this is not necessarily true, because for a large quantum chaotic environment
 it is possible for the rhs of Eq.~(\ref{omega-BW}) to be quite small, such that
 $\omega_{E}$ gets a small value at a given value of $\epsilon$ \cite{pre12-sta}.
 (See Sec.~\ref{sect-big-class-sys}) for more discussions.)

 As another example of comparing the two upper bounds given in Eq.~(\ref{drho-aa}) and Eq.~(\ref{D-REG}),
 one may consider a case, in which the EFs of $|n\ra$ in the uncoupled
 basis have on average a Breit-Wigner form.
 Substituting Eq.~(\ref{wE-BW}) into the rhs of Eq.~(\ref{drho-aa}) for $w_M$,
 one gets the following expression for it,
\begin{gather}\label{}
 \frac{4 \omega_{BW}}{\pi  d_S \Delta} \frac{1}{ \epsilon} + \epsilon,
\end{gather}
 which has a minimum value given by $4\sqrt{ \omega_{BW}/(\pi  d_S \Delta})$.
 This minimum value has the same dependence on $\Delta$ as the rhs of Eq.~(\ref{D-REG}),
 but, the $\lambda$-dependence is different due to that $w_{BW} \propto \lambda^2$.
 Since $\sqrt \lambda$ decreases slower than $\lambda$ with decreasing $\lambda$,
 one notes that
 the upper bound given in Eq.~(\ref{drho-aa}) is smaller than that of Eq.~(\ref{D-REG}) for sufficiently weak interactions.

\section{Offdiagonal elements of $\rho^S$} \label{sect-offdiagRDM-MC}

 In this section, we discuss off-diagonal elements of $\rho^S$.
 In Sec.~\ref{sect-offRDM-gen}, we derive a generic expression for $\rho^{S}_{\alpha \beta} $ with $\alpha \ne \beta$.
 Then, in Sec.~\ref{sect-TLS}, as an illustration of the generic result,
 we discuss a model, in which the central system $S$ is a two-level system
 and the environment $\E$ is a many-body quantum chaotic system.
 Finally, in Sec.~\ref{sect-compare-off-diagonal}, comparisons are given between the obtained results
 and those of Refs.~\cite{RGE12,pre12-sta,YWW-decoh}.

\subsection{A generic expression of $\rho^{S}_{\alpha \beta} $ with $\alpha \ne \beta$}\label{sect-offRDM-gen}

 In this section, for a system $S$ with a nondegenerate spectrum, we derive a generic  expression for
 the  off-diagonal elements
 $\rho^{S}_{\alpha \beta} $,  without any restriction to properties of the $S$-$\E$ interaction. 
 The expression is
\begin{gather}\label{rhoS-rhoSn-ov}
 \rho^S_{\alpha \beta} = \frac{1}{\Delta^S_{\beta\alpha}} Q_{\beta\alpha} \qquad (\alpha \ne \beta), 
\end{gather}
 where $\Delta^S_{\beta\alpha} := e^S_\beta - e^S_\alpha$.
 (See Eq.~(\ref{ovrho}) given below for the definition of $Q_{\beta\alpha}$.)
  Since $\rho^{S0}_{\alpha\beta} =0$, this gives that
\begin{gather}\label{diff-rho-ab}
 |\rho^{S}_{\alpha\beta} - \rho^{S0}_{\alpha\beta}| = \left| \frac{Q_{\beta\alpha}}{\Delta^S_{\beta\alpha}} \right|.
\end{gather}

 To derive Eq.~(\ref{rhoS-rhoSn-ov}), 
 one may start from the following relation implied by the Schr\"{o}dinger Eq.~(\ref{H|n>}),
\begin{gather}\label{ai-H-n}
 \la \alpha i| H^{I} |n\ra =   (E_n - e^S_\alpha -  e_i) C^n_{\alpha i}.
\end{gather}
 Multiplying both sides of Eq.~(\ref{ai-H-n}) by $C^{n*}_{\beta i}$
 and noting that the equality obtained also holds under the exchange of $\alpha \leftrightarrow \beta$,
 one finds that
\begin{subequations}
\begin{gather}
   C^{n*}_{\beta i} \la \alpha i| H^{I} |n\ra =  C^{n*}_{\beta i}  (E_n - e^S_\alpha -  e_i) C^n_{\alpha i},
 \\  C^{n}_{\alpha i} \la n| H^{I} |\beta i\ra =  C^{n}_{\alpha i}  (E_n - e^S_\beta -  e_i) C^{n*}_{\beta i}.
\end{gather}
\end{subequations}
 This gives that
\begin{gather}\label{CC-HI}
  C^{n*}_{\beta i} C^n_{\alpha i} =  \frac{1}{\Delta^S_{\beta\alpha}}
 \left(  C^{n*}_{\beta i} \la \alpha i| H^{I} |n\ra - C^{n}_{\alpha i} \la n| H^{I} |\beta i\ra \right).
\end{gather}
 Substituting Eq.~(\ref{CC-HI}) into Eq.~(\ref{rhoSn}) and writing $C^{n*}_{\beta i} $ and $ C^{n}_{\alpha i}$
 as $\la n|\beta i\ra$ and $\la \alpha i|n\ra$, respectively, one finds that
\begin{gather}\label{rhoSn-[A,HI]}
 \rho^{S(n)}_{\alpha \beta} = \frac{1}{\Delta^S_{\beta\alpha}}  Q^n_{\beta\alpha} \quad (\beta \ne \alpha),
\end{gather}
 where
\begin{gather}\label{Qn-ab}
 Q_{\beta\alpha}^n :=  \la n | [A_{\beta \alpha}, H^{I} ] |n\ra.
\end{gather}
 Here, $A_{\beta \alpha}$ is an operator defined by
\begin{gather}\label{}
 A_{\beta \alpha} := \sum_i  |\beta i \ra  \la \alpha i| = |\beta \ra  \la \alpha | \otimes I^\E.
\end{gather}
 From Eqs.~(\ref{rhoSn-[A,HI]}) and (\ref{rhoS-rhoSn}), it is ready to get Eq.~(\ref{rhoS-rhoSn-ov}),
 with $Q_{\beta\alpha}$ defined as follows,
\begin{gather}\label{ovrho}
 Q_{\beta\alpha}
  := d_{\Gamma }^{-1} \sum_{E_n \in \Gamma} Q_{\beta\alpha}^n.
\end{gather}

 To see more clearly physical meaning of the commutator $[A_{\beta \alpha}, H^{I} ]$,
 let us consider a special case in which $H^I$ has a direct-product form, namely,
\begin{gather}\label{HI-direct-p}
 H^I = H^{IS} \otimes H^{I\E},
\end{gather}
 where $H^{IS}$ and $ H^{I\E}$ are operators acting on the two spaces $\HH^S$ and $\HH^\E$, respectively.
 Elements of  $H^{IS}$ and $H^{I\E}$ in the bases of $|\alpha\ra $ and of $|i\ra$
 are written as
\begin{subequations}\label{HIS-alphabeta}
\begin{gather}
 H^{IS}_{\alpha \beta} \equiv \la\alpha|H^{IS}|\beta\ra,
\\  H^{I\E}_{ij} \equiv \la i | H^{I\E}|j\ra.
\end{gather}
\end{subequations}
 Writing
\begin{gather}\label{}
 H^{IS} =  \sum_{\alpha' \beta'} H^{IS}_{\alpha' \beta'} |\alpha'\ra \la \beta'|,
\end{gather}
 one finds that
\begin{subequations}
\begin{eqnarray}
 |\beta \ra \la \alpha| H^{I} & = &  \sum_{\beta'}  H^{IS}_{\alpha \beta'} |\beta \ra \la \beta'| \otimes H^{I\E},
 \\  H^{I} |\beta \ra \la \alpha| & = & \sum_{\alpha' } H^{IS}_{\alpha' \beta} |\alpha'\ra \la \alpha| \otimes H^{I\E}.
\end{eqnarray}
\end{subequations}
 This gives that
\begin{gather}
 [A_{\beta \alpha}, H^{I} ] =  \sum_{\alpha'}  \left( H^{IS}_{\alpha \alpha'} |\beta \ra \la \alpha'|
 - H^{IS}_{\alpha' \beta} |\alpha'\ra \la \alpha| \right) \otimes H^{I\E},
\end{gather}
 or explicitly,
\begin{gather} \notag
 [A_{\beta \alpha}, H^{I} ] =   (H^{IS}_{\alpha \alpha} - H^{IS}_{\beta \beta} ) |\beta\ra \la \alpha| \otimes H^{I\E}
 \\ +  H^{IS}_{\alpha \beta} ( |\beta \ra \la \beta| -  |\alpha\ra \la \alpha|) \otimes H^{I\E} \notag
 \\ +  \sum_{\alpha' (\ne \alpha, \beta)}  (H^{IS}_{\alpha \alpha'} |\beta \ra \la \alpha'|
 - H^{IS}_{\alpha' \beta} |\alpha'\ra \la \alpha|) \otimes H^{I\E}. \label{[A,HI]-explic}
\end{gather}
 The above expression shows that $[A_{\beta \alpha}, H^{I} ]$ can be regarded as
 certain (non-Hermitian) ``interaction Hamiltonian'' with the system part ``rearranged''.

\subsection{A model with a two-level central system and a quantum chaotic environment}\label{sect-TLS}

 In this section, as an illustration of Eq.~(\ref{rhoS-rhoSn-ov}),
 we discuss a model, in which the subsystem $S$ is a two-level system (a qubit)
 and the environment is a many-body quantum chaotic system
 to which the ETH ansatz \cite{srednicki1999ETH,d2016quantum} is applicable.
 For the simplicity in discussion, we assume that the interaction Hamiltonian $H^I$ has the following properties:
\begin{enumerate}
  \item[(i)] $H^I$ has a direct-product form, as given in Eq.~(\ref{HI-direct-p}),
 with $H^{I\E}$ being a local operator and $H^{IS}_{\alpha\alpha}=0$ for both values of $\alpha$.
  \item[(ii)]  Within the considered energy region, the function $h(e)$, which appears in 
  the ETH ansatz Eq.~(\ref{ETH}) given below, is a constant denote by $h_0$.
\end{enumerate}

 Under the conditions stated above,  Eq.~(\ref{[A,HI]-explic}) gives that
\begin{gather}
 [A_{\beta \alpha}, H^{I} ] =   H^{IS}_{\alpha \beta} ( |\beta \ra \la \beta| -  |\alpha\ra \la \alpha|) \otimes H^{I\E}
\end{gather}
 with $ \beta \ne \alpha$.
 Substituting this result into Eq.~(\ref{Qn-ab}) and making use of the expansion
 of $|n\rangle = \sum_{\alpha i}C_{\alpha i}^n|\alpha i\rangle$, one gets that
\begin{gather}
 Q^{n}_{\alpha \beta} = H^{IS}_{\alpha \beta}
 \sum_{i,j} (C^{n*}_{\beta i} C^{n}_{\beta j} - C^{n*}_{\alpha i} C^{n}_{\alpha j}) H^{I\E}_{ij}. \label{rho-Snab-1}
\end{gather}
 For a local operator $H^{I\E}$, the ETH ansatz predicts that
 \begin{equation}\label{ETH}
 H^{I\E}_{ij} = h(e_i) \delta_{ij} + e^{-S(e_i)/2}g(e_i,e_j)R_{ij},
\end{equation}
 where  ${h}(e)$ is a slowly-varying function of $e$,
 $S(e)$ is proportional to the particle number $N$ of $\E$
 and is related to the microcanonical entropy in a semiclassical treatment,
 $g(e_i,e_j)$ is some smooth function of its variables ($|g|$ being not large),
 and the quantity $R_{ij}$ has certain random feature with
 a normal distribution (zero mean and unit variance).

 Let us compare contributions from the two terms on the rhs of Eq.~(\ref{ETH}) to $Q^{n}_{\alpha \beta}$ in
 Eq.~(\ref{rho-Snab-1}).
 The contribution from the first term is written as
\begin{gather*}
 H^{IS}_{\alpha \beta} \sum_{i} (|C^{n}_{\beta i}|^2 - |C^{n}_{\alpha i}|^2) h(e_i);
\end{gather*}
 meanwhile, that from the second term is 
\begin{gather*}
 H^{IS}_{\alpha \beta} \sum_{i, j} (C^{n*}_{\beta i} C^{n}_{\beta j} - C^{n*}_{\alpha i} C^{n}_{\alpha j}) 
 e^{-S(e_i)/2}g(e_i,e_j)R_{ij}.
\end{gather*}
 When the value of $[h(e_0)\sum_{i} (|C^{n}_{\beta i}|^2 - |C^{n}_{\alpha i}|^2)]$  is not very small, 
 due to the random feature of $R_{ij}$ and the smallness of the term $e^{-S(E)/2}$ at large $N$,
 it is seen that the contribution from the second term is much smaller than that of the first
 term for a sufficiently large environment and, hence, can be neglected.

 Then, making use of Eqs.~(\ref{rhoS-rhoSn}) and (\ref{rhoSn}),
 one gets the following expression of $Q^n_{\alpha \beta}$:
\begin{gather}  \label{Qnab-rhoaa}  
 Q^n_{\beta\alpha} \simeq  h_0  H^{IS}_{\alpha \beta}(\rho^{S(n)}_{\beta \beta}  - \rho^{S(n)}_{\alpha \alpha} ).
\end{gather}
 Substituting Eq.~(\ref{Qnab-rhoaa}) into Eq.~(\ref{rhoSn-[A,HI]}),
 one gets the following simple relation among the elements of the RDM of a single state $|n\ra$,
\begin{gather}\label{rhonS-tls}
 \rho^{S(n)}_{\alpha \beta}  \simeq  \frac{ H^{IS}_{\alpha \beta} h_{0}}{\Delta^S_{\beta\alpha}}
 (\rho^{S(n)}_{\beta \beta}  - \rho^{S(n)}_{\alpha \alpha}) \quad \text{for $\alpha \ne \beta$.}
\end{gather}
 This implies the following relation for elements of $\rho^S$,
\begin{gather}\label{rhoS-tls}
 \rho^S_{\alpha \beta}  \simeq  \frac{ H^{IS}_{\alpha \beta} h_{0}}{\Delta^S_{\beta\alpha}}
 (\rho^{S}_{\beta \beta}  - \rho^{S}_{\alpha \alpha}) \qquad \text{ $(\alpha \ne \beta)$.}
\end{gather}

\subsection{Comparison with results of Refs.~\cite{RGE12,pre12-sta,YWW-decoh}}\label{sect-compare-off-diagonal}

 In this section, we compare results given in the previous two sections
 with those given in Refs.~\cite{RGE12,pre12-sta,YWW-decoh} for off-diagonal elements $\rho^S_{\alpha \beta}$
 with $\alpha \ne \beta$.

 We first discuss Ref.~\cite{pre12-sta}.
 There, only a specific situation was studied for off-diagonal elements $\rho^S_{\alpha\beta}$  (Appendix C
 of Ref.~\cite{pre12-sta}),
 in which quantities like $[H^{IS}_{\alpha \beta} h(e)/ \Delta^S_{\beta\alpha}]$ have very small values.
 It is shown there that the off-diagonal elements $\rho^S_{\alpha\beta}$
 have small values, when the dimension of the effective environmental state space is large.
 This prediction is clearly in agreement with Eq.~(\ref{rhoS-tls}) for a two-level central system.
 For a multi-level system $S$,  agreement can also be found
 by making use of Eq.~(\ref{Qba-multi-S}) to be derived later.

 Next, we compare with  Ref.~\cite{RGE12}.
 It is easy to see that predictions of Eq.~(\ref{diff-rho-ab}) and of Eq.~(\ref{D-REG}) (as Eq.~(2) of Ref.~\cite{RGE12})
 can not always be consistent, because the latter contains a term
 $\Delta^{-1/2}$, while, the former shows no explicit dependence on $\Delta$.
 The difference between the two predictions
 is seen more clearly from Eq.~(\ref{rhoS-tls}) for a two-level system,
 which shows that the value of $|\rho^S_{\alpha \beta} - \rho^{S0}_{\alpha \beta}|$
 does not necessarily decrease with increasing $\Delta $.

 To be precise, let us consider a solvable example, in which
 the interaction Hamiltonian has the simple form of $H^I = H^{IS} \otimes I^\E$ with $[H^{IS}, H^S] \ne 0$.
 Clearly,  one may equivalently  take $\ww H^S =H^S + H^{IS}$ as the self-Hamiltonian of $S$,
 with a vanishing interaction Hamiltonian;
 in other words, the total Hamiltonian $H$ can be reformulated as $H= \ww H^S + H^\E$.
 Under this formulation of $H$,
 following arguments similar to those leading to Eq.~(\ref{rhoS0-solu}), one finds that
 the RDM $\rho^S$ has the following elements in the eigenbasis of $\ww H^S$, denoted by $|\ww \alpha\ra $,
\begin{gather}\label{rhoS-wwaa}
 \rho^{S}_{\ww\alpha \ww\alpha}= \frac{1}{d_{\Gamma^0}}  d^{\E}_{\Gamma\ww\alpha}, \qquad
 \rho^{S}_{\ww\alpha \ww\beta}=0 \ \   (\ww\alpha \ne \ww\beta),
\end{gather}
 where $d^{\E}_{\Gamma\ww\alpha}$ is similar to $d^{\E}_{\Gamma\alpha}$ but related to the state $|\ww\alpha\ra$.
 Transforming from the basis $\{ |\ww \alpha\ra \}$ to $\{ | \alpha\ra \}$,
 since $[H^S, H^{IS}] \ne 0$,  $\rho^S$ usually gets nonzero off-diagonal elements $\rho^S_{\alpha \beta}$
 (unless the values of $d^{\E}_{\Gamma\ww\alpha}$ are independent of the label $\ww \alpha$),
 which do not depend on the value of $\Delta $.

 It is not difficult to check that the above-discussed nonzero $\rho^S_{\alpha \beta}$
 obtained from Eq.~(\ref{rhoS-wwaa})
 are consistent with Eq.~(\ref{rhoS-tls}) related to the formulation of $H= H^S + H^I + H^\E$.
 (See also discussions to be given later in the second part of Sec.~\ref{sect-product-HI-gen}.)
 In contrast, the obtained nonzero $\rho^S_{\alpha \beta}$
 conflict with the prediction of Eq.~(\ref{D-REG}) that they should
 decrease as $\sqrt{1/\Delta}$ or faster with increasing $\Delta$.
 This confliction suggests that Eq.~(\ref{D-REG}) may work under a condition stricter
 than that given in Ref.~\cite{RGE12}.
\footnote{Since the proof of Eq.~(\ref{D-REG}) given in Ref.~\cite{RGE12} is sketchy,
 it is difficult to give a more detailed comparison.}

 Finally, a formula given in Ref.~\cite{YWW-decoh}  for a long-time averaged RDM
 has a form similar to Eq.~(\ref{rhoS-tls}).
 In fact,  that formula of Ref.~\cite{YWW-decoh} can be derived from  Eq.~(\ref{rhoS-tls}) \cite{foot-JZ}.

\section{ Gibbs states with impact of interaction}\label{sect-reno-Gibbs-S}

 In this section, we discuss applications of the generic results given in Eqs.~(\ref{drho-aa}) and (\ref{diff-rho-ab}).
 For brevity, those requirements that have been used in the derivation of these two equations 
 are not to be mentioned below, though they need to be satisfied.

 Specifically, we discuss situations in which the RDM $\rho^S$ may have a Gibbs form,
 besides the well-known case with very weak $S$-$\E$ interactions.
 \footnote{When the interaction is sufficiently weak such that 
 $w_M/d_S \Delta \ll 1$ (with $\epsilon \ll 1$) and $|Q_{\beta\alpha}/\Delta^S_{\beta\alpha}| \ll 1$,
 closeness of $\rho^S$ to $\rho^S_{G}$ is a direct prediction of Eqs.~(\ref{drho-aa}) and (\ref{diff-rho-ab}).
 This is independent of the type of the environment (e.g., integrable or chaotic). }
  In fact, when the interaction is not very weak,
 the values of $(w_M/d_S \Delta)$ and $|Q_{\beta\alpha}/\Delta^S_{\beta\alpha}|$ may be nonnegligible
 and, as a result,  $\rho^S$ may show notable deviation from the Gibbs state $\rho^S_{G}$ in Eq.~(\ref{Gibbs});
 this possibility has already been observed in many numerical simulations (see, e.g., Refs.~\cite{GongJB12,XuDZ14}).
 To study this case analytically, a widely adopted idea is that, instead of $H^S$ used in $\rho^S_{G}$, 
 one may consider a renormalized self-Hamiltonian of the system, which
 takes into account some impact of the interaction  (see, e.g., Ref.~\cite{breuer2002}).

 Below, we show that Eqs.~(\ref{drho-aa}) and (\ref{diff-rho-ab}) supply a generic and reliable framework for 
 realizing the idea mentioned above. 
 Specifically, in Sec.~\ref{sect-RGS-diag}, we give further discussions for renormalized self-Hamiltonian 
 and its usage in Gibbs state.
 Then, in Sec.~\ref{sect-rGibbs-offd}, we show that
 it is possible for $\rho^S$ to be close to a renormalized Gibbs state
 for a big class of (total) systems of physical interest.

\subsection{Renormalized Gibbs state}\label{sect-RGS-diag}

 In this section, we discuss a formulation for Gibbs states with renormalized self-Hamiltonians.
 Within this formulation, a sufficient condition for the closeness of $\rho^S$ to Gibbs state
 can be easily expressed [see Eq.~(\ref{main-cond-ww}) to be given below].

 The possibility of introducing a renormalized self-Hamiltonian for the system $S$ is rooted in the fact that
 the total Hamiltonian in Eq.~(\ref{H}) can always be reformulated in the following way,
\begin{gather}\label{H-ww}
 H  = \ww H^{S} + \ww H^{I} + H^{\E},
\end{gather}
 where
\begin{subequations}
\begin{gather}\label{wwHS}
 \ww H^S = H^S + O^S, 
 \\ \ww H^I = H^I - O^S \otimes I^\E,
\end{gather}
\end{subequations}
 with $O^S$ an operator that acts on the state space of the system $S$.
 The operator $\ww H^S$ can be regarded as a \emph{renormalized self-Hamiltonian} of the system $S$
 and $\ww H^I$ as the corresponding renormalized interaction Hamiltonian.
 We assume that the spectrum of $\ww H^S$ is nondegenerate, too.
 It is not difficult to check that all the generic relations derived in previous sections, 
 particularly Eqs.~(\ref{drho-aa}) and (\ref{diff-rho-ab}), remain valid with this reformulation of the total Hamiltonian.

 Hereafter, we use tilde to indicate items that are obtained  
 under the above-discussed reformulation of the total Hamiltonian, if some change may be caused.
 For example, we use $|\ww \alpha\ra$ to indicate eigenstates of $\ww H^S$.
 While, no tilde is used, if no change may be caused. 
 For example, the RDM $\rho^S$ is independent of the reformulation and,
 hence, there is no need to write a tilde above it;
 similarly, the states $|n\ra$ and $|i\ra$ are also independent of the reformulation.

 Some quantities and relations with tilde are listed below,
\begin{subequations}
\begin{gather}\label{}
 \ww H^S|\ww \alpha\ra = e^S_{\ww \alpha} |\ww \alpha\ra,
 \\ \Delta^S_{\ww\beta \ww\alpha} = e^S_{\ww\beta} - e^S_{\ww\alpha},
 \\ A_{\ww\beta \ww\alpha} = |\ww\beta \ra  \la \ww\alpha | \otimes I^\E,
 \\ \ww Q_{\ww\beta \ww\alpha}^n =  \la n | [A_{\ww\beta \ww\alpha}, \ww H^{I} ] |n\ra. \label{Qn-ab-ww}
\end{gather}
\end{subequations}
 We  use $\ww w_M$ to indicate a width similar to $w_M$,
 but, related to uncoupled states given by $|\ww E_r\ra \equiv |\ww \alpha i\ra$.
 It is not difficult to verify that $\ww\rho^{S0}$,
 the RDM obtained in the case of $\ww H^I =0$, satisfies relations
 similar to those given in Eq.~(\ref{rhoS0-rhoG}), that is, 
\begin{gather}\label{}
 \ww\rho^{S0} \simeq \ww\rho^S_{G},
\end{gather}
 where
\begin{gather}\label{Gibbs-g}
  \ww\rho^S_{G} = e^{-\beta \ww H^S}/ \tr e^{-\beta \ww H^S}.
\end{gather}
 For brevity, we call $\ww\rho^S_{G}$, a Gibbs state with a renormalized self-Hamiltonian,
 \emph{a renormalized Gibbs state}.

 From the reformulated forms of Eqs.~(\ref{drho-aa}) and (\ref{diff-rho-ab}) with tilde, 
 it is easy to find a sufficient condition for $\rho^S \simeq \ww\rho^S_{G}$, as stated below.
\begin{itemize}
  \item If an operator $O^S$ exists, for which the following relations hold with $\epsilon \ll 1$,
\begin{subequations}\label{main-cond-ww}
\begin{gather}\label{main-cond-ww-1}
 \ww w_M/d_S \Delta \ll 1,
 \\ |\ww Q_{\ww\beta \ww\alpha} / \Delta^S_{\ww\beta \ww\alpha}| \ll 1 , \quad \forall \ww\beta \ne \ww\alpha,
 \label{main-cond-ww-2}
\end{gather}
\end{subequations}
 then, $\rho^S \simeq \ww\rho^S_{G}$.
\end{itemize}
 We use $\cs_{\rm sw}$ to indicate the set of operators $O^S$ for which Eq.~(\ref{main-cond-ww-1})
 is satisfied, with ``sw'' standing for ``small width'';
 and  use $\cs_{\rm sQ}$ to indicate the set of operators $O^S$ for which Eq.~(\ref{main-cond-ww-2})
 is satisfied, with ``sQ'' standing for ``small $Q$''.

 For an operator $O^S \in \cs_{\rm sw}$,  
 one has $|\rho^{S}_{\ww\alpha \ww\alpha} - \ww\rho^{S0}_{\ww\alpha \ww\alpha}| \ll 1$
 according to Eq.~(\ref{drho-aa}) with tilde;
 this tells that $\ww\rho^S_{G}$ supplies an appropriate description for diagonal elements of the RDM $\rho^S$
 in the renormalized basis $\{|\ww \alpha\ra\}$.
 In other words, an operator $O^S\in \cs_{\rm sw}$ gives a useful description for the influence of the $S$-$\E$ interaction
 in these diagonal elements of $\rho^S$.
 Meanwhile, for an operator $O^S \in \cs_{\rm sQ}$,  one has
 $|\rho^{S}_{\ww\alpha \ww\beta}| \ll 1$ with $\ww\alpha \ne \ww\beta$ according to Eq.~(\ref{diff-rho-ab}) with tilde;
 this means that the RDM $\rho^S$ is approximately decohered in the eigenbasis of $\ww H^S$. 
 Thus, the eigenbasis of $\ww H^S$ given by $O^S \in \cs_{\rm sQ}$ may supply a statistically preferred basis.
 \footnote{For a further discussion about preferred basis, see the last paragraph of Sec.~\ref{sect-conclusion}.}

 The above-discussed sufficient condition for the closeness of
 $\rho^S$ to $\ww\rho^S_{G}$ can be rewritten as
\begin{gather}\label{cond-Jsw-JsQ-set}
 \cs_{\rm sw} \bigcap \cs_{\rm sQ} \ne \emptyset,
\end{gather}
 where $\emptyset$ indicates the empty set. 
 When the overlap of $\cs_{\rm sw}$ and  $\cs_{\rm sQ}$ is empty, the RDM $\rho^S$ does not
 necessarily have a Gibbs form. 
 In fact,  for a generic total system,
 since the restriction to the operator $O^S$ given in Eq.~(\ref{main-cond-ww-1})
 is quite different from that given in Eq.~(\ref{main-cond-ww-2}),
 there is no reason to expect that the two sets  $\cs_{\rm sw}$ and  $\cs_{\rm sQ}$ 
 must have a nonempty overlap.

 One remark: 
 In some special cases of the total system, the value of $q_1$ in Eq.~(\ref{q1-q0}) may be  evaluated.
 In such a case, one may directly use Eq.~(\ref{drho-simeq}) with tilde, instead of Eq.~(\ref{drho-aa}) with tilde,
 in the above discussions.
 Then, one gets a sufficient and necessary condition for $\rho^S \simeq \ww\rho^S_{G}$,
 which is obtained by simply multiplying the left-hand side of 
 the inequality in Eq.~(\ref{main-cond-ww-1}) by $|q_1|$.

\subsection{Gibbs form of $\rho^S$ for a big class of systems}\label{sect-rGibbs-offd}

 As discussed above, for a generic total system with a nonweak $S$-$\E$ interaction, 
 it is unnecessary for $\rho^S$ to possess a Gibbs form. 
 Physically, of more interest is to study systems with physical restrictions, 
 to see whether $\rho^S$ may possess a Gibbs form.

 In this section, we show that it is possible for $\rho^S$ to be close to $ \ww \rho^S_G$
 for a big class of systems of physical relevance, when the interaction described by $H^I$ is not very weak. 
 We first specify the class of systems in Sec.~\ref{sect-big-class-sys}, 
 next, show validity of Eq.~(\ref{main-cond-ww}) 
 for a direct-product form of $H^I$ in Sec.~\ref{sect-product-HI}, 
 then, discuss a generic form of the interaction Hamiltonian in Sec.~\ref{sect-product-HI-gen},
 and finally, give a simple solvable example and some final remarks in Sec.~\ref{sect-example}.

\subsubsection{A big class of systems}\label{sect-big-class-sys}

 The class of systems to be studied
 includes $S$+$\E$-type systems that satisfy the following three requirements.
\begin{enumerate}
  \item[(i)] The interaction Hamiltonian $H^I$ is local in its environmental part.
  \item[(ii)] The environment is a many-body quantum chaotic system, to which the ETH ansatz is applicable.
  \item[(iii)] There exists an operator $O^S$, denoted by $O^S_{\rm sw}$, for which the EFs of $|n\ra$
 on the basis $\{ |\ww E_r\ra \}$ are narrow,
 in particular,  $\ww w_M \ll \Delta$ and $\ww w_M \ll \min\{ |\Delta^S_{\alpha \beta}|
 \ \text{with} \ \alpha \ne \beta \}$.
\end{enumerate}
 {One notes that narrowness of EFs usually implies narrowness of LDOS.}

 In most physical models, interactions are local. 
 Moreover, although the exact condition under which  the ETH ansatz
 is applicable is still unclear,
 it is expected valid  at least for local operators in many-body quantum chaotic systems
 \cite{d2016quantum,Deutch18}.
 Hence, there are many physical models that satisfy the first two requirements listed above.

 To have a further understanding about the third requirement, we recall a mechanism discussed in Ref.~\cite{pre12-sta},
 by which widths of the total EFs may be considerably reduced, 
 when the environment $\E$ is a many-body quantum chaotic system. 
 There, it is shown that, taking $O^S$ as a partial trace of $H^I$ over certain effective environmental state space,
 an upper bound of the width $\ww w_E$ is proportional to $1/{\Delta_{\cal E}}$,
 where $\Delta_\E$ represents the total energy scale of the environment.
 As a result, with other parameters unchanged, by increasing the size of the environment
 one may get small $\ww w_E$.
 In other words, large size of the chaotic environment may considerably suppress widths of the total EFs
 in the uncoupled basis.

 Based on the above discussions, we conclude that there is a big class of systems 
 that fulfills the above-listed three requirements.

\subsubsection{A direct-product form of local interaction}\label{sect-product-HI}

 In this subsection,  we study the class of systems specified above,
 when the interaction Hamiltonian $H^I$ has the direct-product form in Eq.~(\ref{HI-direct-p}) with a local operator $H^{I\E}$.
 We are to show that Eq.~(\ref{main-cond-ww-2}) is usually valid.
 Together with the property of $\ww w_M \ll \Delta$ which guarantees Eq.~(\ref{main-cond-ww-1}),
 this result implies that $\rho^S \simeq \ww\rho^S_G $,  with $\ww H^S = H^S +O^S_{\rm sw}$.

 To show validity of Eq.~(\ref{main-cond-ww-2}), the key point lies in properties of 
 the quantity $\ww Q_{{\ww \beta}{\ww \alpha}}^n$ defined in Eq.~(\ref{Qn-ab-ww}).
 Before dealing with this quantity, 
 it proves convenient to first study a related quantity $Q_{{\ww \beta}{\ww \alpha}}^n$, defined by 
\begin{gather}\label{Q-ww-ab}
 Q_{\ww\beta \ww\alpha}^n :=  \la n | [A_{\ww\beta \ww\alpha}, H^{I} ] |n\ra.
\end{gather}
 Clearly, the commutator $[A_{\ww\beta \ww\alpha},H^I]$ has an expression similar to Eq.~(\ref{[A,HI]-explic}).
 Substituting this expression into Eq.~(\ref{Q-ww-ab})
 and inserting $\sum_i|i\ra \la i|$, one finds that,
\begin{gather}\notag
 Q_{{\ww \beta}{\ww \alpha}}^n = \sum_{ij} \Big\{
 (H^{IS}_{{\ww \alpha} {\ww \alpha}} - H^{IS}_{{\ww \beta} {\ww \beta}} )  \la n |{\ww \beta} i\ra 
 \la {\ww \alpha} j|n\ra  H^{I\E}_{ij}
 \\ +  H^{IS}_{{\ww \alpha} {\ww \beta}} ( \la n|{\ww \beta} i\ra \la {\ww \beta} j|n\ra -  \la n|{\ww \alpha} i\ra 
 \la {\ww \alpha} j|n\ra)  H^{I\E}_{ij}  \label{Qba-S-multi}
 \\ +  \sum_{{\ww \alpha}' (\ne {\ww \alpha}, {\ww \beta})}  (H^{IS}_{{\ww \alpha} {\ww \alpha}'} \la n
 |{\ww \beta} i\ra \la {\ww \alpha}' j|n\ra
 - H^{IS}_{{\ww \alpha}' {\ww \beta}} \la n|{\ww \alpha}' i\ra \la {\ww \alpha} j|n\ra ) H^{I\E}_{ij} \notag
 \Big\}.  
\end{gather}

 The rhs of Eq.~(\ref{Qba-S-multi}) can be simplified.
 To this end, we note that the main-body region of the EF of $|n\ra$ in the basis $\{ |\ww E_r\ra\}$,
 which contains all its significant components, lies within the 
 energy region of $[E_n -\ww w_M, E_n + \ww w_M]$ [cf.~Eq.~(\ref{EF-region})].
 In the equivalent $|\ww \alpha i\ra$ form of the basis,
 this implies the following approximate expression of $|n\ra$,
\begin{gather}\label{n-narrow-Upsilon}
 |n\ra \simeq \sum_{\ww \alpha} \sum_{e_i \in \Upsilon_{\ww \alpha}^n} C^n_{{\ww \alpha} i} |{\ww \alpha} i\ra,
\end{gather}
 where $\Upsilon_{\ww \alpha}^n$ indicates the environmental energy region
 of $[E_n - e^S_{\ww \alpha} - \ww w_M, E_n - e^S_{\ww \alpha} + \ww w_M]$.
 The smallness of $\ww w_M$ implies narrowness of each region $\Upsilon_{\ww \alpha}^n$.
 As a result, for $e_i \in \Upsilon_{\ww \alpha}^n$, one has $e_i \simeq E_n - e^S_{\ww \alpha}$;
 and, since $\ww w_M \ll \min\{ |\Delta^S_{\alpha \beta}|\}$, 
 there is no overlap between  $\Upsilon_{\ww \alpha}^n$ of different ${\ww \alpha}$.

 To see influences of the above-discussed properties of the EFs in the quantity $Q_{{\ww \beta}{\ww \alpha}}^n$,
 as an example,  let us consider the first part of the rhs of Eq.~(\ref{Qba-S-multi}), 
 namely $\sum_{ij} H^{IS}_{{\ww \alpha} {\ww \alpha}}  \la n |{\ww \beta} i\ra \la {\ww \alpha} j|n\ra  H^{I\E}_{ij}$,
 which we denote by $W_1$.
 Substituting Eq.~(\ref{n-narrow-Upsilon}) into $W_1$, one finds that
\begin{gather}\label{W1}
 W_1 \simeq  H^{IS}_{{\ww \alpha} {\ww \alpha}}  \sum_{e_i \in \Upsilon_{\ww \beta}^n} 
 \sum_{e_j \in \Upsilon_{\ww \alpha}^n} 
 C^{n*}_{{\ww \beta} i} C^n_{{\ww \alpha} j}  H^{I\E}_{ij},  \quad \ww\alpha \ne \ww\beta.
\end{gather}
 Note that the term $H^{I\E}_{ij}$ is already given in Eq.~(\ref{ETH}) by the ETH ansatz. 
 Since $\Upsilon_{\ww \alpha}^n$ and $\Upsilon_{\ww \beta}^n$ have no overlap, 
 the first part on the rhs of Eq.~(\ref{ETH}) gives negligible contribution to $W_1$. 
 Furthermore, we note that 
\begin{gather}\label{}
 \Big | \sum_{e_i \in \Upsilon_{\ww \beta}^n} 
 \sum_{e_j \in \Upsilon_{\ww \alpha}^n}  C^{n*}_{{\ww \beta} i} C^n_{{\ww \alpha} j} R_{ij} \Big | < 1;
\end{gather}
 hence, due to the term $e^{-S/2}$, 
 the second part on the rhs of Eq.~(\ref{ETH}) also gives negligible contribution to $W_1$.

 Other parts of the rhs of Eq.~(\ref{Qba-S-multi}) can be discussed in a similar way.
 It turns out that nonnegligible contributions to $Q_{{\ww \beta}{\ww \alpha}}^n$ 
 come from the middle line of Eq.~(\ref{Qba-S-multi}), with $H^{I\E}_{ij}$
 given by the first part on the rhs of Eq.~(\ref{ETH}).
 The result is
\begin{gather} \notag
 Q_{{\ww \beta}{\ww \alpha}}^n \simeq \sum_{i}
 H^{IS}_{{\ww \alpha} {\ww \beta}} \Big( |\la n|{\ww \beta} i\ra|^2 -  |\la n|{\ww \alpha} i\ra|^2 \Big) h(e_i)
 \\ \simeq H^{IS}_{{\ww \alpha} {\ww \beta}}  \Big(   h(E_n-e^S_{\ww \beta}) \rho^{S(n)}_{{\ww \beta} {\ww \beta}} -
 h(E_n-e^S_{\ww \alpha}) \rho^{S(n)}_{{\ww \alpha} {\ww \alpha}} \Big), \label{Qba-multi-S}
\end{gather}
 where $\rho^{S(n)}_{{\ww \alpha} {\ww \alpha}}$ is defined in Eq.~(\ref{rhoSn})
 and slow variation of the function $h(e)$ has been used in the derivation  of the second equality.

 Now, we discuss the quantity $\ww Q_{{\ww \beta}{\ww \alpha}}^n$.
 Let us consider the difference $\Delta Q_{{\ww \beta}{\ww \alpha}}^n
 := \ww Q_{{\ww \beta}{\ww \alpha}}^n - Q_{{\ww \beta}{\ww \alpha}}^n$.
 Clearly, it has the following expression, 
\begin{gather}\label{}
 \Delta Q_{{\ww \beta}{\ww \alpha}}^n
 = - \la n | [A_{\ww\beta \ww\alpha}, O^S_{\rm sw}\otimes I^\E ] |n\ra,
\end{gather}
 which has a form similar to $Q_{{\ww \beta}{\ww \alpha}}^n$ in Eq.~(\ref{Q-ww-ab}). 
 Hence, this difference can be written in a form like Eq.~(\ref{Qba-S-multi}), 
 but, with $H^{IS}$ replaced by $(-O^S_{\rm sw})$ and $H^{I\E}$ by $I^\E$.
 Then, making use of the fact that $I^{\E}_{ij} \equiv \la i|I^\E|j\ra = \delta_{ij}$, 
 similar to Eq.~(\ref{Qba-multi-S}), one gets that
\begin{gather} 
 \Delta Q_{{\ww \beta}{\ww \alpha}}^n 
 \simeq (O^S_{\rm sw})_{{\ww \alpha} {\ww \beta}}  \Big( 
  \rho^{S(n)}_{{\ww \alpha} {\ww \alpha}} - \rho^{S(n)}_{{\ww \beta} {\ww \beta}}  \Big), \label{Qba-multi-S-delta}
\end{gather}
 where $(O^S_{\rm sw})_{{\ww \alpha} {\ww \beta}} = \la \ww\alpha | O^S_{\rm sw} | \ww \beta \ra$.
 This gives that
\begin{gather} \notag
 \ww Q_{{\ww \beta}{\ww \alpha}}^n 
  \simeq \Big( H^{IS}_{{\ww \alpha} {\ww \beta}}  h(E_n-e^S_{\ww \beta}) - (O^S_{\rm sw})_{{\ww \alpha} {\ww \beta}}
 \Big)  \rho^{S(n)}_{{\ww \beta} {\ww \beta}} 
  \\ - \Big(H^{IS}_{{\ww \alpha} {\ww \beta}} h(E_n-e^S_{\ww \alpha}) 
  - (O^S_{\rm sw})_{{\ww \alpha} {\ww \beta}} \Big)
  \rho^{S(n)}_{{\ww \alpha} {\ww \alpha}}. \label{ww-Qba-multi-S}
\end{gather}

 To go further, we note that the assumed smallness of $\ww w_M$ 
 should require some restriction to properties of the interaction.
 To see this point clearly, let us consider an arbitrary pair of basis states,
 say, $|{\ww \alpha} i\ra$ and $|{\ww \beta} i\ra$,
 with  the same label $i$ and ${\ww \alpha} \ne {\ww \beta}$.
 The level spacing of these two states is given by $\Delta^S_{{\ww \beta}{\ww \alpha}}$.
 Making use of the ETH ansatz, one finds that the coupling between these two states,
 i.e., the off-diagonal element $\la {\ww \alpha} i|\ww H^I|{\ww \beta} i\ra$,  is written as
\begin{gather}\label{HI-elem}
 \la {\ww \alpha} i|\ww H^I|{\ww \beta} i\ra \simeq H^{IS}_{{\ww \alpha} {\ww \beta}} h(e_i)
 - (O^S_{\rm sw})_{{\ww \alpha} {\ww \beta}}.
\end{gather}
 According to the perturbation theory,  the assumed narrowness of the EF of $|n\ra$ usually requires
 that $| \la {\ww \alpha} i|\ww H^I|{\ww \beta} i\ra  / \Delta^S_{{\ww \beta}{\ww \alpha}}| \ll 1$,
 at least for those levels $e_i$ that are not far from the values of $(E_n -e^S_{{\ww \alpha} ({\ww \beta})})$.
 Then, making use of Eq.~(\ref{ww-Qba-multi-S}), one finds that
 $|\ww Q_{{\ww \beta}{\ww \alpha}}^n / \Delta^S_{{\ww \beta}{\ww \alpha}}| \ll 1$.
 Therefore, Eq.~(\ref{main-cond-ww-2}) usually holds. 

\subsubsection{A generic form of local interaction}\label{sect-product-HI-gen}

 In this subsection, for the class of systems discussed above, we show that Eq.~(\ref{main-cond-ww-2}) also holds
 under a generic local interaction Hamiltonian, implying that $\rho^S \simeq \ww\rho^S_G $.
 
 A generic $H^I$ can always be written as a sum of direct-product terms, i.e., 
\begin{equation}\label{HI-gen}
H^I = \sum_\nu  H^{I,\nu},  
\end{equation}
 where
\begin{gather}\label{}
 H^{I,\nu} = \sum_\nu H^{IS,\nu}\otimes H^{I\E,\nu}.
\end{gather}
 Locality of the interaction implies that all the operators $H^{I\E,\nu}$ are local operators.
 The ETH ansatz is applicable to each of them. 
 We assume that the number of the terms $H^{I,\nu}$, namely $\sum_\nu 1$, is not large.

 To study this generic case, we write $Q_{\ww\beta \ww\alpha}^{n}$ as 
 $Q_{\ww\beta \ww\alpha}^{n} =\sum_\nu Q_{\ww\beta \ww\alpha}^{n,\nu}$, where
\begin{gather}\label{Qn-wwab-nu}
 Q_{\ww\beta\ww\alpha}^{n,\nu} =  \la n | [A_{\ww\beta \ww\alpha}, H^{I, \nu} ] |n\ra.
\end{gather}
 It is easy to see that the quantity $Q_{\ww\beta\ww\alpha}^{n,\nu}$ can be treated in exactly the same
 way as done in the previous subsection for $Q_{\ww\beta\ww\alpha}^{n}$ and,
 as a result, similar to Eq.~(\ref{Qba-multi-S}), it is written as
\begin{gather} 
 Q_{{\ww \beta}{\ww \alpha}}^{n,\nu}  
 \simeq H^{IS,\nu}_{{\ww \alpha} {\ww \beta}}  \Big(   h^\nu(E_n-e^S_{\ww \beta}) \rho^{S(n)}_{{\ww \beta} {\ww \beta}} -
 h^\nu(E_n-e^S_{\ww \alpha}) \rho^{S(n)}_{{\ww \alpha} {\ww \alpha}} \Big), \label{Qba-multi-S-nu}
\end{gather}
 where $h^\nu(e)$ indicates a function that appears when the ETH ansatz  is applied 
 to the operator $H^{I\E,\nu}$ [cf.~Eq.~(\ref{ETH})].
 Then, noting that Eq.~(\ref{Qba-multi-S-delta}) is still valid, one finds that
\begin{gather} \notag
 \ww Q_{{\ww \beta}{\ww \alpha}}^n   \simeq  
   \Big((O^S_{\rm sw})_{{\ww \alpha} {\ww \beta}} - \sum_\nu 
  H^{IS,\nu}_{{\ww \alpha} {\ww \beta}} h^\nu(E_n-e^S_{\ww \alpha})  \Big)
  \rho^{S(n)}_{{\ww \alpha} {\ww \alpha}} , 
  \\ -\Big( (O^S_{\rm sw})_{{\ww \alpha} {\ww \beta}} 
  -\sum_\nu H^{IS,\nu}_{{\ww \alpha} {\ww \beta}}  h^\nu(E_n-e^S_{\ww \beta}) 
 \Big)  \rho^{S(n)}_{{\ww \beta} {\ww \beta}}.
  \label{ww-Qba-multi-S-gen}
\end{gather}

 For the same reason as that discussed in the previous subsection, 
 the assumed narrowness of the EF of $|n\ra$ usually requires
 that $| \la {\ww \alpha} i|\ww H^I|{\ww \beta} i\ra  / \Delta^S_{{\ww \beta}{\ww \alpha}}| \ll 1$.
 It is easy to check that the elements $\la {\ww \alpha} i|\ww H^I|{\ww \beta} i\ra$ are now written as
\begin{gather}\label{HI-elem-gen}
 \la {\ww \alpha} i|\ww H^I|{\ww \beta} i\ra \simeq  - (O^S_{\rm sw})_{{\ww \alpha} {\ww \beta}}
 + \sum_\nu H^{IS,\nu}_{{\ww \alpha} {\ww \beta}} h^\nu(e_i).
\end{gather}
 Then, it is ready to see that
$|\ww Q_{{\ww \beta}{\ww \alpha}}^n / \Delta^S_{{\ww \beta}{\ww \alpha}}| \ll 1$.
 Therefore, usually Eq.~(\ref{main-cond-ww-2}) holds, too. 
 
 \subsubsection{One example and some concluding remarks}\label{sect-example}
 
 In this subsection, as an example, we consider the model discussed in Sec.~\ref{sect-TLS}
 and explicitly show closeness of $\rho^S$ to $ \ww\rho^S_G $ under nonweak interactions. 
 We also give some concluding remarks.

 In the model discussed in Sec.~\ref{sect-TLS}, 
 making use of Eq.~(\ref{rhoS-tls}), it is straightforward to verify that $\rho^S$
 has the following matrix form in the eigenbasis of $H^S$,
\begin{gather}\label{matrix-S-qubit}
 \left[ \rho^S \right] \simeq \frac{\rho^{S}_{\beta \beta}  - \rho^{S}_{\alpha \alpha}}{\Delta^S_{\beta\alpha}}
 \left(   \begin{array}{cc}      0 & H^{IS}_{\alpha \beta} h_{0} \\
       H^{IS}_{\beta \alpha} h_{0} & \Delta^S_{\beta\alpha} \\
     \end{array}    \right) + \rho^S_{\alpha\alpha} \left(   \begin{array}{cc}      1 & 0  \\
       0 & 1 \\    \end{array}    \right).
\end{gather}
 The main result of Ref.~\cite{pre12-sta} is applicable to this model,
 which shows that $\ww w_E$ is small for a sufficiently large environment, 
 with a renormalized self-Hamiltonian $\ww H^S = H^S +h_0 H^{IS}$;
 this implies that $O^S_{\rm sw} = h_0 H^{IS}$.
 Hence, this model belongs to the big class of systems discussed above.

 In the basis of $|\alpha\ra$, $\ww H^S$ has the following matrix form,
\begin{gather}\label{ww-HS-pre12}
 \left[ \ww H^S \right] = \left(   \begin{array}{cc}
       e^S_\alpha & H^{IS}_{\alpha \beta} h_{0} \\
       H^{IS}_{\beta \alpha} h_{0} & e^S_\beta \\
     \end{array}    \right).
\end{gather}
 From the above expressions, it is seen that $\left[ \rho^S \right]$ and $\left[ \ww H^S \right]$
 are diagonalized by almost a same transformation.
 Then, it is straightforward to check that Eq.~(\ref{main-cond-ww}) is satisfied and $\rho^S \simeq \ww \rho^S_G$.

 Finally, we give some remarks on properties of a many-body quantum chaotic system with local interactions,
 which is described by an MC ensemble.
 Let us consider a division of this system into a small part $S$ and a large part $\E$.
 This $S$+$\E$ configuration of the system belongs to the class of systems discussed previously,
 if $\E$ is a sufficiently large quantum chaotic system
 and if the mechanism discussed in Ref.~\cite{pre12-sta} for narrowness of EFs works. 
 Then, according to discussions given previously, even with nonweak interactions, 
 all such small parts $S$ are described by (renormalized) Gibbs states with a same temperature,
 which is determined by  the density of states of the total system.

\section{RDM computed from typical states} \label{sect-RDM-typical}

 In this section, we discuss the RDM of $S$, which is computed from a typical
 state of the total system in the energy shell $\Gamma$.
 We denote it by $\rho^S_{\rm ty}$.
 In particular, we derive estimates to the elements
 $(\rho^S_{\rm ty})_{\alpha \beta} \equiv \la \alpha |\rho^S_{\rm ty}|\beta\ra$,
 which may enable one to get details of the difference between $\rho^S_{\rm ty}$ and
 the previously discussed RDM $\rho^S$.
 For brevity, we use $O (x)$ to indicate an undetermined quantity,
 whose order of magnitude is the same as that of a quantity $x$.

\subsection{Basic properties of $\rho^S_{\rm ty} $}

 We use $|\Psi_{\rm ty}^{\Gamma }\ra$ to denote a normalized typical vector in the subspace $\HH_{\Gamma }$,
 written as
\begin{equation}\label{aPt-de}
 |\Psi_{\rm ty}^{\Gamma }\ra = \N_{\Gamma }^{-1} \sum_{E_n \in \Gamma } D_n |n\ra ,
\end{equation}
 where the real and imaginary parts of $D_n$ are independent Gaussian random
 variables, with mean zero and variance $1/2$, and $\N_{\Gamma }$ is the normalization coefficient.
 The RDM $\rho^S_{\rm ty}$ is given by
 \be \rho^S_{\rm ty} = \tr_{\E} \left( |\Psi_{\rm ty}^{\Gamma }\ra \la \Psi_{\rm ty}^{\Gamma }| \right).
 \ee
 As shown in  Ref.~\cite{PSW06}, the averaged trace distance between $\rho^S$ and  $\rho^S_{\rm ty} $ satisfies
\begin{gather}\label{}
 \la D( \rho^S, \rho^S_{\rm ty}) \ra \le \frac 12 \sqrt{\frac{d_S^2}{d_{\Gamma}}}.
\end{gather}

 To study $\rho^S_{\rm ty} $, we expand the typical state $|\Psi_{\rm ty}^{\Gamma }\ra$
 according to the system $S$'s states $|\alpha\ra$, i.e.,
 \begin{equation}\label{Psi-ty-alpha}
 |\Psi_{\rm ty}^{\Gamma }\ra = \N_{\Gamma }^{-1} \sum_{\alpha} |\alpha\ra |\Phi^\E_\alpha\ra,
 \end{equation}
 where
 \begin{eqnarray} \label{Omega}
 |\Phi^\E_\alpha\ra =  \sum_{i} \left(\sum_{E_n \in \Gamma }
 D_n  C^n_{\alpha i} \right) | i\ra .
 \end{eqnarray}
 It is easy to verify that
\begin{gather}\label{}
 (\rho^S_{\rm ty})_{\alpha \beta}  = \N_{\Gamma }^{-2} \la \Phi^\E_\beta |\Phi^\E_\alpha \ra,
 \\   \la \Phi^\E_\beta |\Phi^\E_\alpha \ra   = \sum_i  \sum_{E_n\in \Gamma }
 \sum_{E_{n'}\in \Gamma }  D^{*}_n D_{n'}  C^{n*}_{\beta i}   C^{n'}_{\alpha i}.  \label{Osum}
\end{gather}
 From Eq.~(\ref{aPt-de}) and the randomness of the coefficients $D_n$, one finds that
\begin{gather}\label{norN}
 \N_{\Gamma }^2 = \sum_{E_n \in \Gamma} |D_n|^2
 = d_{\Gamma} + O(\sqrt{d_{\Gamma}}),
\end{gather}
 and, as a result,
 \begin{eqnarray} \label{rsab}
 (\rho^S_{\rm ty})_{\alpha \beta} = \frac{1}{d_{\Gamma}}  \left( 1+ O\big( \frac{1}{\sqrt{d_{\Gamma}}} \big) \right)
 \la \Phi^\E_\beta |\Phi^\E_\alpha \ra .
 \end{eqnarray}

 In the two sections  following this one, we discuss properties of the diagonal part $\la \Phi^\E_\alpha|\Phi^\E_\alpha\ra $
 and of the off-diagonal part  $\la \Phi^\E_\beta|\Phi^\E_\alpha\ra $ with $\alpha \ne \beta$, separately.
 There, it proves convenient to divide the overlap $ \la \Phi^\E_\beta|\Phi^\E_\alpha\ra$  into two parts,
\begin{gather}\label{Omg-K}
 \la \Phi^\E_\beta|\Phi^\E_\alpha\ra = K^{(1)}_{\beta \alpha} + K^{(2)}_{\beta \alpha },
\end{gather}
 where $K^{(1)}_{\beta \alpha }$ represents the diagonal contribution of the rhs of Eq.~(\ref{Osum}) with $n=n'$
 and $K^{(2)}_{\beta\alpha }$ is for the off-diagonal contribution with $n \ne n'$.

\subsection{Diagonal overlap --- $\la \Phi^\E_\alpha|\Phi^\E_\alpha\ra $}\label{sect-K1}

 In this section, we derive estimates to the two quantities $ K^{(1)}_{\alpha \alpha} $
 and $ K^{(2)}_{\alpha \alpha} $, from which an estimate to $\la \Phi^\E_\alpha|\Phi^\E_\alpha\ra $
 can be gotten from Eq.~(\ref{Omg-K}).
 
\subsubsection{An estimate to $ K^{(1)}_{\alpha \alpha} $}

 By definition, the quantity $ K^{(1)}_{\alpha \alpha} $ is written as
\begin{gather}\label{K1-aa}
 K^{(1)}_{\alpha \alpha} = \sum_{E_n\in \Gamma }  |D_{n}|^2 B_{\alpha n},
\end{gather}
 where
\begin{gather}\label{B-na}
  B_{\alpha n}= \sum_i |C^n_{\alpha i}|^2.
\end{gather}
 We use $ \ov{B}_\alpha$ to indicate the average value of $B_{\alpha n}$ within the energy shell $\Gamma$, i.e.,
\begin{gather}\label{}
 \ov{B}_\alpha := \frac{1}{d_\Gamma} \sum_{E_n \in \Gamma} B_{\alpha n}.
\end{gather}
 It is straightforward to verify that the RDM element $\rho^S_{\alpha \alpha}$ [see Eq.~(\ref{rhoS-rhoSn})]
 is equal to $ \ov{B}_\alpha$, i.e.,
\begin{gather}\label{}
 \rho^S_{\alpha \alpha} = \ov{B}_\alpha .
\end{gather}
 Since the average of $|D_n|^2$ is equal to $1$, the sum of $ \sum_{E_n\in \Gamma } (|D_{n}|^2-1) B_{\alpha n}$
 has an absolute value that has the same order of magnitude as $ \rho^S_{\alpha \alpha} \sqrt{d_{\Gamma}}$.
 Then, from Eq.~(\ref{K1-aa}) one gets that
\begin{gather}\label{K1a-1}
  K^{(1)}_{\alpha \alpha} = d_\Gamma \rho^S_{\alpha \alpha}
  + \rho^S_{\alpha \alpha} e^{i\varphi_1} O( \sqrt{d_{\Gamma}}),
\end{gather}
 where $\varphi_1$ is equal to either $0$ or $\pi$ in a random way.
 Note that $\rho^S_{\alpha \alpha}$ are elements of $\rho^S$ discussed in previous sections.

\subsubsection{An estimate to $ K^{(2)}_{\alpha \alpha} $}

 The quantity $ K^{(2)}_{\alpha \alpha} $ is defined by
\begin{gather}\label{K1-J}
 K^{(2)}_{\alpha \alpha} = \sum_i J_{\alpha i},
\end{gather}
 where
\begin{gather}
 J_{\alpha i} =  \sum_{E_n\in \Gamma } \ \ \sum_{ E_{n'} (\ne E_n) \in \Gamma }
 D^{*}_n D_{n'} C^{n*}_{\alpha i} C^{n'}_{\alpha i}. \label{J2ai}
\end{gather}
 Below, we show that 
\begin{gather}\notag
 K^{(2)}_{\alpha \alpha} =  \sqrt{N_{2}^{\E\alpha}}  e^{i\theta_2} O(1)
 + (\sigma_{1} e^{i\theta_1} +  \sigma_3 e^{i\theta_3}) \sqrt{ \frac{ 2w_L d^\E_{\Gamma \alpha}}{\Delta}  } O(1)
 \\ -  a_2 \epsilon \sqrt{N_{2}^{\E\alpha}} e^{i\theta_2} O(1) + b_0 \epsilon \sqrt{d_\Gamma} e^{i\theta_0} O(1),
 \label{K2a-2}
\end{gather}
 where $\theta_{0,1,2,3}=0$ or $\pi$  in a random way, and $\sigma_{1,3}$ and $b_0$ are 
 undetermined parameters satisfying $0< \sigma_{1,3},b_0 <1$.
 Equation (\ref{K2a-2}) shows that,
 for a large $d_\Gamma$, usually $K^{(2)}_{\alpha \alpha} $ gives a small contribution to
 $(\rho^S_{\rm ty})_{\alpha \alpha}$ in Eq.~(\ref{rsab}).

 To derive Eq.~(\ref{K2a-2}), 
 we divide $K^{(2)}_{\alpha \alpha}$ into subparts according to the regions $\R_{\kappa}^{\E\alpha}$,
 like what was done in Sec.~\ref{sect-MC-DRM-diagonal}, that is,
\begin{gather}\label{K2-K2k}
 K^{(2)}_{\alpha \alpha} = \sum_{\kappa =0}^3 K^{(2)}_{\alpha\alpha, \kappa},
\end{gather}
 where
\begin{gather}\label{K2k-J}
 K^{(2)}_{\alpha\alpha, \kappa} = \sum_{e_i\in \R_{\kappa}^{\E\alpha}} J_{\alpha i}.
\end{gather}
 It proves convenient to introduce the following quantity,
\begin{gather}\label{}
  {\cal I}_{\alpha i} =  \sum_{E_n\in \Gamma }   |D_{n}|^2 |C^n_{\alpha i}|^2. \label{J1ai}
\end{gather}
 We note that, due to the randomness of the components $D_n$,
 usually, $J_{\alpha i}$ and $ {\cal I}_{\alpha i}$ have the following relation,
\begin{gather}\label{J2-J1}
 J_{\alpha i} =  {\cal I}_{\alpha i} e^{i\vartheta_i}  O(1),
\end{gather}
 with $\vartheta_i =0$ or $\pi$ in a random way.

 First, we discuss the term $ K^{(2)}_{\alpha\alpha, 2}$, the contribution
 coming from the central region  $\R_{2}^{\E\alpha}$.
 For $e_i$ lying in this region, like $I_{\alpha i}$ in Eq.~(\ref{Iai}), ${\cal I}_{\alpha i}$ also fluctuate around
 $(1- a_2 \epsilon)$.
 Then, making use of Eq.~(\ref{J2-J1}), one finds that
\begin{gather}\label{}
 K^{(2)}_{\alpha\alpha, 2} = \sqrt{N_{2}^{\E\alpha}}   (1- a_2 \epsilon) e^{i\theta_2} O(1),
\end{gather}
 with $\theta_2=0$ or $\pi$ in a random way.

 Second, we discuss $ K^{(2)}_{\alpha\alpha, 1}$, coming from the region $\R_{1}^{\E\alpha}$.
 We use $i_0$ and $i_f$ to indicate the starting and ending labels of $\R_{1}^{\E\alpha}$.
 In this region, ${\cal I}_{\alpha i}$ is close $1$ for $e_i$ close to the region $\R_{2}^{\E\alpha}$,
 while, it is small for $e_i$ close to the region $\R_{0}^{\E\alpha}$.
 Loosely speaking, with the label $i$ increasing from $i_0$ to $i_f$, ${\cal I}_{\alpha i}$ increases
 on average from some value close to $0$ to some value close to $1$.
 It is this difference in the values of ${\cal I}_{\alpha i}$ that makes it uneasy to
 get an estimate to $ K^{(2)}_{\alpha\alpha, 1}$.

 To circumvent the above-mentioned difficult,
 we construct new variables from $J_{\alpha i}$.
 At the first step, we construct a series of variables, denoted by $X_s^{(1)}$
 with $s=0,1\ldots, s_f$, where $s_f$ is given by the integer part of $(i_f-i_0)/2$.
 Specifically,
\begin{gather}\label{Xs1}
 X_s^{(1)} = J_{\alpha (i_0+s)} + J_{\alpha (i_f-s)}  \ \ \text{for} \  s=0,1\ldots, s_f-1;
\end{gather}
 $X_{s_f}^{(1)}$ is given by Eq.~(\ref{Xs1}) if $(i_f-i_0)$ is odd, otherwise,
 $X_{s_f}^{(1)} = J_{\alpha (i_0+s_f)}$.
 Clearly, the variance of $ X_s^{(1)}$ is equal to the sum of the variances of
 $J_{\alpha (i_0+s)}$ and $J_{\alpha (i_f-s)}$ for $s\le  s_f-1$.
 We proceed following the above procedure, until an $L$-th step is reached,
 at which most $X_t^{(L)}$ have similar variances.
 It is easy to see that
\begin{gather}\label{}
  K^{(2)}_{\alpha \alpha,1} = \sum_t X_t^{(L)}.
\end{gather}
 We assume that $N_{1}^{\E\alpha}$ is sufficiently large, such that $N_{1}^{\E\alpha} \gg 2^L$.
 Note that the number of the variables $X_t^{(L)}$ at the $L$-the step is about $N_{1}^{\E\alpha}/2^L$.

 According to the construction of $X_t^{(L)}$,
 the sum of the variances of $X_t^{(L)}$ over $t$  is equal to that of $J_{\alpha i}$
 over $i$ with $e_i \in K^{(2)}_{\alpha\alpha, 1}$.
 This implies that the averaged variance of $X_t^{(L)}$ is around $(\sigma_1^2 2^L)$,
 where $\sigma_1^2$ is the averaged variance of these $J_{\alpha i}$.
 Then, one finds that
\begin{gather}\label{}
 K^{(2)}_{\alpha \alpha,1}
 =  \sigma_1 e^{i\theta_1} \sqrt{N_{1}^{\E\alpha}} O(1),
\end{gather}
 with $\theta_1=0$ or $\pi$ in a random way.
 The above arguments are also applicable to $K^{(2)}_{\alpha \alpha,3} $.
 Thus, when the environmental density of states does not change much around the energy shell $\Gamma_0$,
 we find that
\begin{gather}
 K^{(2)}_{\alpha\alpha , \kappa} =  \sigma_{\kappa}  e^{i\theta_\kappa}  \sqrt{ \frac{ 2w_L d^\E_{\Gamma \alpha}}{\Delta}  }
 O(1) \quad \text{for} \ \kappa =1,3.
\end{gather}
 It is easy to check that $0< \sigma_{1,3} <1$.

 Third, we discuss $ K^{(2)}_{\alpha\alpha, 0} $, coming from
 $e_i$ lying in the region $\R_{0}^{\E\alpha}$.
 Making use of Eq.~(\ref{J2-J1}), we write it in the following form,
\begin{gather}
 K^{(2)}_{\alpha\alpha, 0} = \sum_{E_n \in \Gamma} |D_n|^2
 \left( \sum_{e_i\in \R_{0}^{\E\alpha}} |C^n_{\alpha i}|^2 x_i \right),
\end{gather}
 where $x_i$ represents a  random variable, whose variance is of the order of magnitude of $1$.
 Noting Eq.~(\ref{cna2<ep}), one sees that
\begin{gather}
 \left| \sum_{e_i\in \R_{0}^{\E\alpha}}   |C^n_{\alpha i}|^2 x_i \right|  =  b_0 \epsilon ,
\end{gather}
 where $b_0$ is some undetermined parameter satisfying $0< b_0<1$
 (usually $b_0 \ll 1$).
 Then, one gets that
\begin{gather}
 K^{(2)}_{\alpha\alpha, 0} = b_0 \epsilon \sqrt{d_\Gamma} e^{i\theta_0} O(1),
\end{gather}
 with $\theta_0=0$ or $\pi$ in a random way.
 Summarizing the above results, one finally gets Eq.~(\ref{K2a-2}).

\subsection{Offdiagonal terms}

 For an off-diagonal term $\la \Phi^\E_\beta|\Phi^\E_\alpha\ra$ with $\alpha \ne \beta$,  its two parts
 are written in the following forms,
\begin{gather}\label{K1ab-J}
 K^{(1)}_{\beta\alpha }  =  \sum_{E_n\in \Gamma } \sum_i    |D_{n}|^2 C^{n*}_{\beta i} C^{n}_{\alpha i},
 \\ \label{K2ab-J}
 K^{(2)}_{\beta\alpha }  =  \sum_{E_n\in \Gamma } \ \ \sum_{ E_{n'} (\ne E_n) \in \Gamma } \sum_i
 D^{*}_n D_{n'} C^{n*}_{\beta i} C^{n'}_{\alpha i}.
\end{gather}
 The two quantities $K^{(1)}_{\beta\alpha }$ and $ K^{(2)}_{\beta\alpha }$ can be studied by
 a method similar to that used in the previous section for diagonal terms,
 and qualitatively similar results can be obtained.

 For example, to study $K^{(1)}_{\beta\alpha }$, we write it as
\begin{gather} \label{J1abi}
 K^{(1)}_{\beta\alpha }  = \sum_{E_n\in \Gamma } |D_{n}|^2 \rho^n_{\alpha \beta}.
\end{gather}
 Similar to $K^{(1)}_{\alpha\alpha}$ discussed previously, one finds that
\begin{gather}\label{}
 K^{(1)}_{\beta\alpha } = d_\Gamma \rho^S_{\beta\alpha} + \sqrt{d_\Gamma} e^{i\theta_{\beta\alpha}}
 O(\sigma_{\beta\alpha}),
\end{gather}
 where $\theta_{\beta\alpha}$ is some undetermined phase and $\sigma_{\beta\alpha}^2$
 represents the variance of $\rho^n_{\beta \alpha} $.

 Finally, substituting results thus obtained for the off-diagonal contributions $\la \Phi^\E_\beta|\Phi^\E_\alpha\ra$  and
 the previously-obtained results for diagonal elements in Eqs.~(\ref{K1a-1}) and (\ref{K2a-2}) into Eq.~(\ref{rsab}),
 it is straightforward to get an estimate to $[\rho^S_{\alpha\beta} - (\rho^S_{\rm ty})_{\alpha\beta}]$.

\section{Conclusions and discussions}\label{sect-conclusion}

 Main results of this paper consist of two parts.
 The first part supplies a generic framework for the study of closeness of 
 the RDM $\rho^S$ of a generic interacting small system to the Gibbs state.
 The second part contains a sufficient condition for the above-mentioned closeness and,
 as an application, a study of a big class of systems of physical relevance.

 In the first part, we consider a generic, isolated, and large quantum system, 
 which is described by an MC ensemble;
 it is divided into a generic small subsystem $S$, with a RDM $\rho^S$, and an environment $\E$.
 There are two restrictions to the systems: (i) The system $S$ has a nondegenerate spectrum;  and (ii) the 
 $S$-$\E$ interaction is not very strong, such that it has negligible influence in the density of states of the total system.
 We have studied the difference between the elements of $\rho^{S}$
 in the eigenbasis of the subsystem's self-Hamiltonian and the corresponding elements of $\rho^{S0}$, 
 the latter of which is the DRM obtained in the case of none $S$-$\E$ interaction.

 Specifically, upper bounds have been derived for the difference between diagonal elements, 
 which are mainly confined by the ratio of the maximum width of the total EFs in the
 uncoupled basis to the width of the MC energy shell.
 Meanwhile, the difference between off-diagonal elements has been found 
 given by the ratio of certain property of the interaction Hamiltonian to the related level spacing of the
 system $S$.
 These results show that, although $\rho^S$ and the Gibbs state
 are close under sufficiently weak $S$-$\E$ interactions, notable deviation may appear 
 under interactions not very weak.
 Besides, the difference has also been studied between $\rho^S$ and 
 the RDM that is computed from a typical state of the total system.

 The second part contains applications of generic results of the first part,  
 for $S$-$\E$ interactions that are not necessarily weak.
 A sufficient condition is given, under which $\rho^S$ is close to a Gibbs state
 that contains a renormalized self-Hamiltonian of $S$; the renormalization is due to impact of the interaction.
 For a big class of total systems of physical relevance, 
 $\rho^S$ are shown to be usually close to (renormalized) Gibbs states. 
 In this class, a total system contains a subsystem $S$ that is locally coupled to an environment as
 a many-body quantum chaotic system, to which the ETH ansatz is applicable.

 The above-discussed results may be useful in the study of several important 
 topics of current interest, such as thermalization,  quantum thermodynamics, and decoherence.
 As an example, the possibility of closeness of $\rho^S$ to a (renormalized) Gibbs state under nonweak $S$-$\E$ interactions
 should be useful in the study of some thermalization processes. 
 Particularly, this is because, if such a Gibbs state may exist and be found,  it may represent a steady state of 
 some thermalization process. 
 Moreover, this closeness and the expression of the renormalized self-Hamiltonian 
 of $S$ (if found) may supply important clues,
 when deriving a master equation for a related process,

 As another example, the results may also find applications in the study of 
 an important concept in the field of decoherence, i.e., the so-called preferred (pointer) basis
 \cite{Zurek81,zurek2003decoherence,Schloss05,JZKGKS03,wiseman2010quantum,PZ99,DK00,
 wang2008entanglement}.
 For example, suppose that a renormalized Gibbs state supplies an appropriate description for
 a steady state of a thermalization process with decoherence.
 Then, the eigenbasis of the related renormalized self-Hamiltonian may 
 give a (statistically) preferred basis \cite{foot-pre14-ps}.

\acknowledgements

 The author is grateful to Qingchen Li, Jiaozi Wang, and Yan Gu for valuable discussions and suggestions.
 This work was partially supported by the Natural Science Foundation of China under Grants
 No.~11535011 and No.~11775210.

\end{document}